\documentclass[nopreprint,numbers]{jasatex}

\usepackage{graphicx}
\usepackage{dcolumn}
\usepackage{amsmath,amsfonts}
\usepackage{bm}

\usepackage{epsfig} 
\usepackage{subfigure}

\usepackage{epstopdf}

\begin{document}

\preprint{AIP/123-QED}

\title[Hegewisch et al.: Constructing acoustic timefronts...]
{Constructing acoustic timefronts using random matrix theory}
\author{Katherine C.~Hegewisch}
\author{Steven Tomsovic}
\affiliation{Department of Physics and Astronomy, Washington State University, Pullman, WA~~99164-2814}

\date{\today}%

\begin{abstract}

In a recent letter [Europhys.~Lett.~{\bf 97}, 34002 (2012)], random matrix theory is introduced for long-range acoustic propagation in the ocean.  The theory is expressed in terms of unitary propagation matrices that represent the scattering between acoustic modes due to sound speed fluctuations induced by the ocean's internal waves.  The scattering exhibits a power-law decay as a function of the differences in mode numbers thereby generating a power-law, banded, random unitary matrix ensemble.  This work gives a more complete account of that approach and extends the methods to the construction of an ensemble of acoustic timefronts.   The result is a very efficient method for studying the statistical properties of timefronts at various propagation ranges that agrees well with propagation based on the parabolic equation.  It helps identify which information about the ocean environment can be deduced from the timefronts and how to connect features of the data to that environmental information.  It also makes direct connections to methods used in other disordered wave guide contexts where the use of random matrix theory has a multi-decade history.

\end{abstract}
\maketitle

\section{\label{sec:level1} Introduction}

Our interest in this paper is in long-range acoustic propagation that is vertically confined by the deep-ocean sound channel.   In addition to the vertical wave guide confinement, there is important multiple scattering of the acoustic waves that is induced by local fluctuations in the ocean environment~\cite{Flatte79,Munk95}.   A number of previous experiments were designed with an acoustic source placed centrally in the wave guide, i.e.~near the so-called sound channel axis, and receivers placed in vertical arrays~\cite{Worcester99,Colosi99}.  The acoustic waves arrive at the receivers in timefronts that feature early arriving isolated branches that give way to complicated finales with the majority of the energy.  The timefronts have both structural and statistical properties which are the focus of this investigation.

An improved understanding of the acoustic timefronts would lead to the possibility of deducing new information about the state of the ocean.  However, the multiple scattering greatly complicates the analysis of experimental data, and in particular is responsible for a partial loss of information about the ocean~\cite{Makarov09,Simmen97,Brown03,Beronvera03,Tomsovic10}.  It is still not fully resolved as to what information can be deduced from long-range propagation and how to extract it.

Since the 1970's, numerical studies have often utilized paraxial optical approximations~\cite{Tappert77} coupled with Fourier transform based projections to simulate acoustic timefronts.  It has advanced over the years into an indispensable computational tool, but it is not well suited to extracting analytic relationships between properties of the ocean and timefronts or providing insights that come readily with ray methods.  It furthermore is still computationally intensive enough that performing larger scale ensembles of calculations for statistical studies would be prohibitive.  

Although, the ocean wave guide has its own unique features, such as the timefront itself, it also possesses common characteristics with many other wave guide problems.  Indeed, long-range ocean acoustics has long been recognized as a context in which quantum/wave chaos is present~\cite{Palmer88, Smith92a,Smith92b}.  Therefore it has strong parallels with physical systems such as mesoscopic disordered and ballistic electronic conductors~\cite{Beenakker97RMP,Marcus92} and quantum point contacts~\cite{Topinka00}.  For example, it was noted that quantum point contacts channel wave energy, called coherent branching in that context (not to be confused with branches of a timefront), through the behavior of its underlying rays~\cite{Topinka01} in much the same way as occurs in long-range ocean acoustics~\cite{Wolfson01}.  Curiously, one of the key theoretical tools of quantum chaos, random matrix theory, was only recently initiated by us in the ocean acoustic  propagation context~\cite{Hegewisch12,Hegewisch10}.  Partly this is due to the fact that traditional random matrix theory does not straightforwardly translate there.  In other acoustic contexts, e.g.~certain problems in elasto-dynamics~\cite{Weaver89}, there exists a more direct application in which the featureless ensembles of Wigner and Dyson~\cite{Wigner55,Dyson62e} apply without modification.  In ocean acoustics, random matrix theory enters in terms of how amplitudes in the acoustic modes mix with propagation in range, but some dynamical information must be incorporated that is absent in the standard ensembles.  The mixing is captured by the unitary propagation matrix of complex probability amplitudes for modal transition~\cite{Dozier78,Dozier78b,Morozov07,Colosi09}.

Random matrix theory is connected to a maximum entropy hypothesis~\cite{Mello10} and is well suited for investigating what information can be deduced and how it appears in long-range acoustic propagation.   Thus, there is a multi-fold purpose to this paper; i) to give a more complete description of how random matrix theory can be introduced into long-range acoustic propagation; ii) apply it for the first time to the full construction of timefronts; iii) investigate possible new connections between environmental information and experimentally accessible quantities; and iv)  examine some of the statistical properties that emerge from this approach.

This paper is organized as follows.  The next section gives basic background material both for the ocean acoustics system and random matrix theory.  This serves to provide the essential physical context, introduce possibly unfamiliar concepts from random matrix theory, specify what will be taken into account, and define notation.  Section~\ref{construction} gives the derivation of the random matrix ensembles to be used to model the acoustical system and points out which environmental information is accounted for in this first application and construction.  Section~\ref{results} shows the comparison between timefronts produced with parabolic equation calculations and random matrix ensembles.  Both average and fluctuating properties are considered.  We finish with a discussion of the merits of the random matrix approach and future possibilities.

\section{Background - Ocean acoustics and random matrix theory}
\label{background}

The following four subsections specify the features of the long-range acoustics system which are taken into account in this study.  For our purposes, the idea is not to have the greatest realism and most accurate modeling of the ocean, but rather to have enough complexity to illustrate how random matrix theory can be applied and be helpful.  The final subsection contains some elementary background in random matrix theory since this may be unfamiliar to many readers.  We summarize a few useful concepts such as ergodicity, statistical measures, and information.

\subsection{Parabolic equation}
The parabolic equation for the propagation of an acoustic wavefield $\Psi(z,r;k)$ as a continuous 
wave~\cite{Tappert77} with wavenumber $k$, depth $z$, and range $r$ through 
a sound speed potential $V$ is
\begin{equation}
\label{eq:parabolic}
      \frac{i}{k}\frac{\partial}{\partial r}\Psi(z,r;k) = -\frac{1}{2k^2}\frac{\partial^2}{\partial z^2}\Psi(z,r;k) +V(z,r) \Psi(z,r;k)
\end{equation}
\noindent This equation assumes that the propagation is essentially forward and thus the energy moves  within a narrow band of angles relative to the sound channel axis and that there is no significant backscattering. 

The sound speed can be decomposed into the reference sound speed, $c_0$, and fluctuations, $\delta c$, about the reference: $c(z,r) = c_0 + \delta c(z,r)$ with $\delta c(z,r) << c_0$, so the potential is approximated as follows: 
\begin{equation}
\label{potential}
V(z,r)=\frac{1}{2}\left(1-\left(\frac{c_0}{c(z,r)}
\right)^2\right)  \approx \frac{\delta c(z,r)}{c_0} \ .  
\end{equation}
Our calculations use $c_0=1.49$ km/sec.
\subsection{Sound speed model}

The model potential $V(z,r)$ takes the form
\begin{equation}
\label{eq:V}
V(z,r) = \frac{\delta c(z,r)}{c_0} = V_0(z) + \epsilon V_1(z,r) \ ,
\end{equation}
\noindent where $V_0$ represents the relative change of the sound speed due to the
vertical waveguide confinement and $\epsilon V_1$ represents the relative sound speed fluctuations due to internal waves, which are small perturbations on the wave guide~\cite{Brown03}; $\epsilon$ is of order $10^{-3}$.    In fact, other perturbations could easily be accounted for and added into $\epsilon V_1(z,r)$, but for the purposes of this work accounting for the internal waves is sufficient and the overall physical context no less general.

We take the vertical confinement to be expressed by Munk's canonical model~\cite{Munk74}
\begin{equation}
\label{eq:Munks}
V_0(z)= \frac{B\gamma}{2}
\left[e^{-\eta(z)} - 1+\eta(z)\right] \ ,
\end{equation}
\noindent where $\eta(z) =2[z-z_a]/B$,
$z_a=1$ km  is the sound channel axis, $B$ is the thermocline depth scale
giving the approximate width of the sound channel, and $\gamma$ is a
constant representing the overall strength of the confinement.  

The fluctuations due to internal waves are captured with the statistical ensemble model 
developed by Colosi and Brown~\cite{Colosi98}, which is a sum over internal wave modes $j$ with horizontal wavenumbers $k_l$.  It has the form
\begin{eqnarray}
\label{eq:colosibrown_c}
\epsilon V_1(z,r)&=& \sum_{j=1}^{j_{max}} \sum_{l=1}^{l_{max}} V_j(z;k_l) \cos\left(\phi_{jl} +k_l r
\right) \ ,
\end{eqnarray}
where $V_j(z;k_l)$ contains the depth dependence and weighting of the $j^{th}$ internal wave mode, and there is a set of uniformly chosen random numbers $\{\phi_{jl}\}$ on the interval $[0;2\pi)$ that create a random wave field $\epsilon V_1(z,r)$.    The upper limits on the summations are set by computational, frequency, and environmental considerations~\cite{Hegewisch05}.  The weighting $V_j(z;k_l)$ decays with increasing $j$ and beyond a maximum value makes negligible contributions.  For the horizontal wavenumber summation over $l$, there is a minimum and maximum relevant horizontal wavenumber dependent upon $f_0$ and thus $l_{max}$ is chosen so that there is a sufficiently dense uniformly selected values of $k_l$ included in the internal wave field construction.  See Appendix~\ref{ap:soundspeed} for further details, the definition of $V_j(z;k_l)$ (Eq.~(\ref{a2})), and the specification of the $k_l$ included in the summation. 

\subsection{Acoustic modes and unitary propagation}
\begin{figure}
\includegraphics[width=3.5 in]{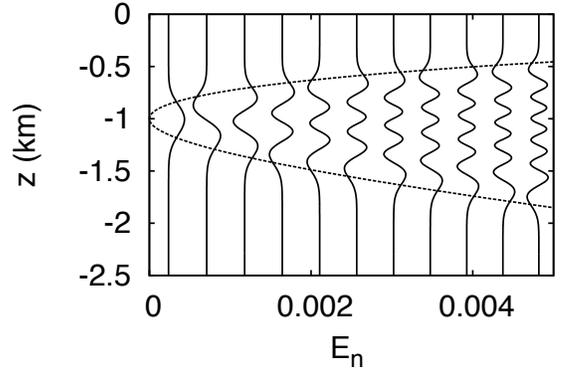}
\caption{Eigenmodes, $\psi_n(z;k)$, of the waveguide potential, $V_0$, are
plotted versus ocean depth, $z$ for $n=0,1,...,10$ at their eigenvalues $E_n$. 
The potential $V_0$ is plotted as a dashed line.}
\label{fig:modes}
\end{figure}

\noindent For a range-independent potential, $V_0(z)$, a general solution to the parabolic equation
can be written in terms of separable solutions
\begin{eqnarray}
\label{propagation}
\Psi(z,r;k) &=& \sum_m a_m(k) e^{-i k r E_m} \psi_m(z;k) \ ,
\end{eqnarray}
\noindent where $\psi_m(z;k)$ satisfies the Sturm-Liouville eigenvalue problem,
\begin{eqnarray}
\label{eq:Sturm}
-\frac{1}{2}\frac{d^2 \psi_m}{dz^2} +k^2 V_0(z) \psi_m &=& k^2 E_m \psi_m 
\end{eqnarray}
\noindent with $\psi_m(z;k)$ and $E_m$ the eigenfunctions (modes) and eigenvalues, respectively.  The boundary condition for Eq.~(\ref{eq:Sturm}) is that the eigenmodes vanish sufficiently rapidly as $|z|\rightarrow \infty$, the eigenmodes are square integrable, and the phase is chosen such that the solutions are real.  The set of weightings $\{ a_m(k) \}$ determine the initial state ($r=0$).  The eigenvalues give each mode a horizontal wavenumber $kE_n$ for their propagation.  Figure~\ref{fig:modes} illustrates the depth dependence of a few of the  first modes.  Note that slow, adiabatic changes with range can be accounted for as well in exactly the same manner.  There is no loss in generality by considering the vertical confinement to be $V_0(z)$.

The modes are orthonormal and serve as a complete basis for arbitrary wavefields in depth.  The scattering between modes has been analyzed by looking at both the range evolution of the coupling coefficients~\cite{Dozier78} and of the mode amplitudes~\cite{Voronovich06, Colosi09,Voronovich09}; the former representing scattering between mode m and mode n  during the propagation from internal wave modes.  Information about the acoustic scattering in each component of the modal basis is found by propagating each mode $\psi_n(z;k)$ as a continuous wave using the full parabolic equation Eq.~(\ref{eq:parabolic}), denoted as $\Psi_n(z,r;k)$, and projecting onto the set of $\psi_m(z;k)$.  Mathematically, the unitary propagation matrix elements $U_{mn}(r;k)$ are 
\begin{eqnarray}
\label{eq:Cmn}
U_{mn}(r;k) &= &\int \Psi_n(z,r;k) \psi^*_m(z;k)  dz \ . 
\end{eqnarray}
These elements represent the probability amplitude of making a transition from mode $n$ to mode $m$.  Since propagation with the parabolic equation preserves the norm and the orthonormality of two initially orthonormal wave fields,  this matrix must be a unitary matrix. Therefore $U U^\dagger = I$, where $U^\dagger$ is the Hermitian conjugate (the complex conjugate transpose of matrix $U$) and $I$ is the identity matrix.    If propagated in the unperturbed system with potential $V_0(z)$ (i.e.~internal waves are turned off, $\epsilon V_1=0$), each propagated mode $\Psi_n(z,r;k)$ accumulates only a phase, i.e.~$\Psi_n(z,r;k)=e^{-ikrE_n}\psi_n(z;k)$, and the unitary matrix is diagonal.  If propagated using the internal wave perturbation, each propagated mode $\Psi_n(z,r;k)$ mixes into other modes.  The amount  and nature of the mixing is a consequence of the properties of the internal wave field. 

\subsection{Acoustic timefront}

Source signals that are localized in time, as opposed to being continuous waves, generate signals at the downrange receivers as a series of pulses with time.  Considering these signals for all depths forms a structure called the acoustic timefront.  The construction of the timefronts requires a weighted Fourier transfrom of the solutions at fixed-$k$.  A Gaussian-shaped, localized pulse in time is used here, which results from a Gaussian weight in the Fourier transform.   More specifically, the complex acoustic timefront $\Phi$ at a range $r$ is constructed from a superposition of  the continuous wave arrivals $\Psi(z,r;k)$ at each source wavenumber~\cite{Tappert77}, 
\begin{equation}
\Phi(z,r;t) =\frac{1}{\sqrt{2 \pi r \sigma_k^2}} \int  \Psi(z,r;k) e^{-\frac{(k-k_0)^2}{2 \sigma_k^2} - i k c_0 \left(t-\frac{r}{c_0}\right) }{\rm d}k \ , 
\label{eq:timefront}
\end{equation}
\noindent where $k_0 = 2 \pi f_0/c_0$ is the central source wavenumber and $\sigma_k$ is the standard deviation.  A source frequency of $f_0=75$ Hz with a standard deviation $18.75$ Hz is used to model a source with a 3 dB bandwidth of 37.5 Hz~\cite{Hegewisch05}. 

The range-dependent wavefield can be written in terms of the propagated modes as $\Psi(z,r;k) = \sum_{n} a_n(k) \Psi_n(z,r;k)$.  Thus, the timefront in terms of modes is computed as
\begin{eqnarray} 
\label{tf}
\Phi(z,r;t) &=& \frac{1}{\sqrt{2 \pi \sigma_k^2 r }}\sum_{m,n}  \int {\rm d}k\  a_n(k) U_{mn}(r;k) \times \nonumber \\  
&& \psi_m(z;k) e^{ - i k c_0\left(  t - \frac{r}{c_0} \right) - \frac{(k-k_0)^2}{2 \sigma_k^2}} \ .
\end{eqnarray}
This is the experimentally measurable quantity from which to deduce information about the internal waves and sound speed profile.
\begin{figure}
\centering
     \subfigure{
          \includegraphics[width= 3.3 in]{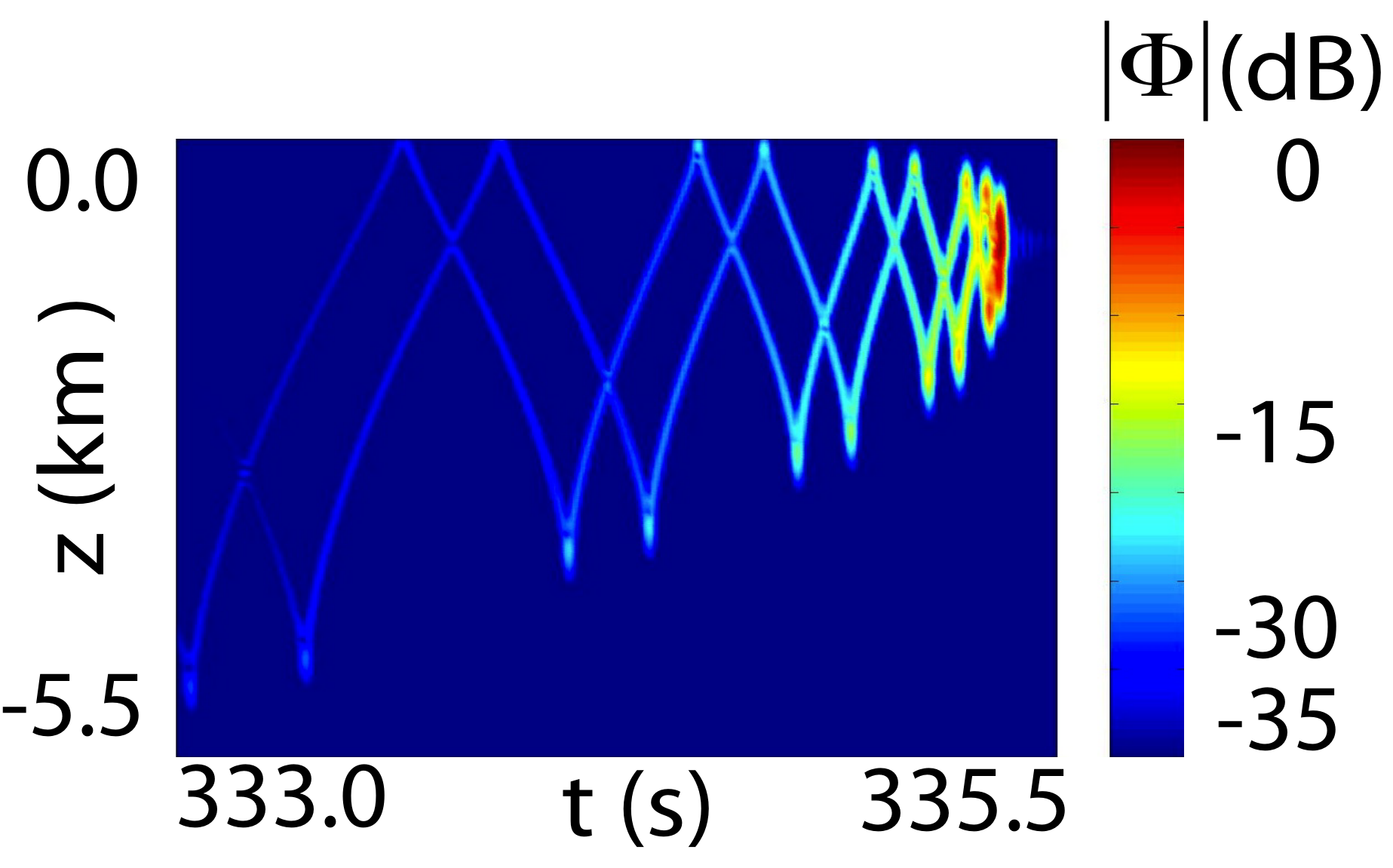}} 
     \subfigure{
          \includegraphics[width= 3.3 in]{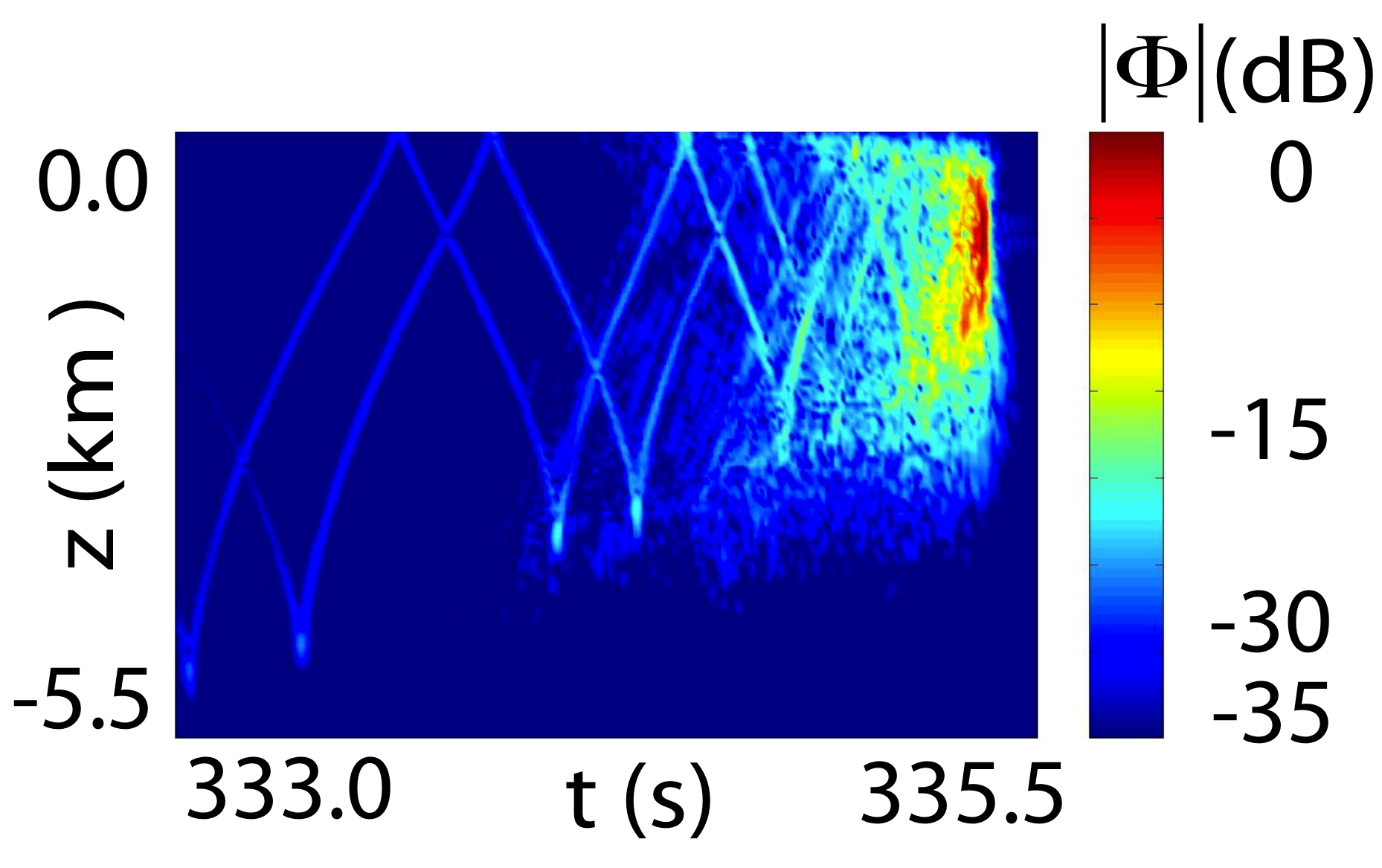}}
\caption{Color density plots of timefronts  calculated using the parabolic equation and weighted Fourier transform, which illustrate their basic features for propagation of $1000$ km.  The upper figure is without including internal wave scattering, the lower figure with internal waves.}
\label{fig:tf}
\end{figure}
Figure \ref{fig:tf} shows timefronts propagated with and without internal wave scattering.  Without scattering, it has well defined, stable branches throughout its entire duration.  With scattering, there are branches with fluctuating intensities and small shifts in locations in the early arriving part.  Most of the power is contained in the latest part of the arrival where the intensities still fluctuate, but the branches overlap and disappear into a complicated finale.  The complicated structure is deterministic since for a fixed perturbation, the resulting propagation is unique, but over the ensemble of internal wave realizations, it can be analyzed statistically.

\subsection{Random matrix theory}

Random matrix theory gives a means to understand the statistical properties of wave chaotic systems.  As such, it can be thought of as a branch of statistical mechanics; see Ref.~~\onlinecite{Beenakker97RMP} and references therein for a review of applications to wave guide problems and quantum transport in mesoscopic conductors.  A complete theory begins with a construction of a random matrix ensemble that models the chaotic system of interest sufficiently well so as to reproduce all of its statistical properties.  Ideally, that would be followed by analytic calculations as expectation values over the ensemble which would allow one to predict and describe the system's statistical properties without even actually having to construct the ensemble numerically.  Often such calculations are quite sophisticated and one does not know whether they are tractable in advance.  Even in their absence, the ensemble may be quite useful for modeling purposes and for understanding what information exists in the system's statistical properties.  The original ensembles, introduced by Wigner and Dyson~\cite{Wigner55,Dyson62e}, are featureless and contain no information about the system other than what fundamental symmetries are preserved, i.e.~time-reversal symmetry, angular momentum~\cite{PorterBook}.  For long-range ocean acoustics, some information must be built into the ensembles, although one does not know, a priori, which information and how to incorporate it.  That process begins in the next section.

A critical concept is that random matrix ensembles possess a generalized ergodic property.  This crudely means that in the limit of increasing matrix dimensionality, each individual system's statistical properties match those of the ensemble itself.  For the standard ensembles of Wigner and Dyson, this was proven by Pandey~\cite{Pandey79}.  This is a stronger feature than is needed for the ocean acoustics system as repeated measurements can be taken hours apart by which time the precise state of the ocean changes significantly enough so as to be able to generate roughly ``independent measurements.''  Nevertheless, even a single acoustic field propagation is statistically likely to possess the structure and fluctuation properties (sampled properly within the timefront) inherent in an ensemble of such independent measurements, after careful accounting for diurnal and seasonal secular behaviors.

In order to apply random matrix theory fully, it is necessary to be precise about the statistical measures to be calculated.  For a timefront, many quantities have been discussed such as the wander in the branches from one realization to the next, the statistical fluctuations in amplitudes in the finale or somewhere on a particular branch, the spreading within a branch, and any bias in the timefront to be advanced or retarded relative to the perturbation free case~\cite{Munk95}.  It is of interest to understand the frequency, range, and perturbation strength dependencies of the statistical measures.  As we are not attempting here analytical ensemble calculations, we do not give exact statistical measure definitions for this study, but rather just recognize that it would be necessary to begin such calculations.

Finally, information is another critical concept.  In the standard ensembles, only information about fundamental symmetries is included.  As an example, if the system is time reversal invariant, the random Hermitian matrix representing the system Hamiltonian has real Gaussian random entries for the matrix elements.  For systems in which time-reversal is strongly broken, the matrix elements are complex Gaussian random entries.  This is insufficient for ocean acoustics as at least some limited information about the nature of the perturbation has to be taken into account.  A method for constructing appropriate random matrix ensembles is given in the next section.

\section{Random matrix ensemble construction}
\label{construction}
The unitary propagation matrix of Eq.~(\ref{eq:Cmn}) is the most suitable object for which to construct a random matrix ensemble.  Our construction~\cite{Hegewisch12}, here given in greater detail, begins with the property of multiplicative unitary propagation in which the product of two unitary propagators gives the propagator for the sum of their ranges.  The idea is to determine a range of propagation, $r_b$, over which the matrix $U_j(r_b;k)$ initially gains certain statistical properties, specified ahead.  The $U_j(r_b;k)$ becomes a building block propagator~\cite{Perez07} in which ideally each is statistically independent of its predecessors and successors.  The block is quite reminiscent of the slices used in the Chernov method~\cite{Chernov75}.  The subscript $j$ indicates then that this unitary propagator is one particular realization of the building block random ensemble.  A single member of the full random matrix ensemble is given by the product of $N$ statistically, independently-drawn, building blocks as
\begin{eqnarray}
\label{eq:product}
U_{rmt}(Nr_b;k)= \prod_{j=1}^N U_j(r_b;k) \ , 
\end{eqnarray}
\noindent where the full range of propagation is given by $Nr_b$.  It remains to be determined what an optimal range for $r_b$ is and what are the statistical laws governing the matrix elements of $U_j(r_b;k)$.  The subscript $rmt$ is added to distinguish this ensemble member from the unitary propagator that results from solving the parabolic equation for a specific potential realization.  In the case that the total range of propagation is just $r_b$, then the full random matrix ensemble reduces to the individual building block random matrix ensemble.  

Ideally, the propagation range $r_b$ is as short as possible subject to the constraint that the physical system generates $\{U_j(r_b;k)\}$ with the statistical properties: i) correlations between $U_j(r_b;k)$ and $U_{j+m}(r_b;k)$ are small for any nonzero integer $m$, ii) the phases of each matrix element $\left[U_j(r_b;k)\right]_{mn}$ are largely randomized, and iii) dynamical correlations between neighboring matrix elements within a given $U_j(r_b;k)$ are minimal.  All the matrix elements of $U_j(r_b;k)$ are correlated due to the constraints of unitarity.  However, the dynamical correlations just mentioned are in addition to those.  If $r_b$ is too short in range, there is little scattering, $U_j(r_b;k)$ is nearly diagonal, and the matrix elements are highly correlated with seemingly little randomness.  In a quantum/wave chaotic system, as propagation range increases, these statistical properties do appear.  On the other hand, if $r_b$ is too long, one may miss universal properties that emerge from the product structure of the propagation.  Just as products of independent random variables generate lognormally distributed quantities due to a central limit theorem (under certain conditions), products of independently chosen structured random matrices may also lead to universal laws.  In addition, for too long a range, there would be no way to attempt using range-dependent perturbation theory to guide the ensemble construction.

For very short propagation ranges, perturbation theory gives an approximation for the propagator as
\begin{equation}
U(r;k) \approx \Lambda(r;k) \left( I - i\epsilon k\int^r_0 {\rm d}r^\prime V_I (r^\prime,k) \right) \ ,
\label{pert1}
\end{equation}
where the elements of the diagonal matrix $\Lambda$ are given by
\begin{equation}
\label{lambda}
\Lambda_{mn}(r;k)=\delta_{mn}\exp(-ik E_m r)\ .
\end{equation}
and $V_I(r^\prime,k)$ is the operator corresponding to $V_1(z,r)$ in the interaction picture and whose modal matrix elements are given by
\begin{equation}
\label{vimn}
\epsilon \left[V_I (r^\prime;k)\right]_{mn} = {\rm e}^{ik(E_m-E_n)r^\prime} \int {\rm d}z \psi^*_m(z;k) \psi_n(z;k) V_1(z,r^\prime) \ .
\end{equation}
This is similar to the coupling matrix of Ref.~~\onlinecite{Dozier78}.  $\Lambda(r;k)$ is the unitary propagator if the perturbation is absent and only the wave guide potential taken into account.  To include adiabatic range dependence would require that the argument of the exponential in $\Lambda_{mn}(r;k)$ be an integral over range as $E_m$ would vary slowly and to incorporate the $r^\prime$-dependence of $\left[E_m,E_n, \psi_m(z;k), \psi_n(z;k) \right]$ in $V_I (r^\prime;k)$.

First order perturbation theory does not maintain unitarity.  However, there are multiple ways to create unitary propagators, which can be connected to the perturbation theory result.  One convenient approach is to apply the Cayley transformation such that as $\epsilon\rightarrow 0 $, the diagonal result is recovered.  For any Hermitian operator $A$ and identity matrix $I$, 
\begin{eqnarray}
\label{eq:unitary}
U =\Lambda(I+i\epsilon A)^{-1}(I-i\epsilon A) \ , 
\end{eqnarray}
$U$ is guaranteed to be unitary.  The matrix elements of $\Lambda$ are given in Eq.~(\ref{lambda}).  The key is to Taylor series expand the inverse operator and connect the statistical properties of the $A$ matrix elements with the statistical properties of the matrix elements of $V_I$ by using Eq.~(\ref{pert1}).  This gives
\begin{equation}
\label{av}
A= \frac{k}{2} \int^{r_b}_0 {\rm d}r^\prime V_I(r^\prime;k) \ ,
\end{equation}
if the expansions of $U$ and $U(r_b;k)$ are equated.  Actually, for the construction of the random matrix ensemble, one does not equate $A$ to the right hand side of Eq.~(\ref{av}), one equates the probability density of each $A$ matrix element in the random matrix ensemble to the probability density of a matrix element of the right hand side evaluated over the ensemble of internal wave fields.  As long as $\epsilon$ is small, this construction restores unitarity while altering very little the precise fluctuation properties of the individual matrix elements predicted by the integral over $V_I(r;k)$.

For the ocean's internal waves, the perturbation is much stronger near the surface than deep in the ocean to such an extent that an approximation is sometimes invoked in which there is an abrupt perturbation near the surface and everywhere else it vanishes.  In that approximation, each cycle of the wave  has a single dominant independent perturbation.  This approximation is not needed, but it leads to the idea of equating $r_b$ roughly with the cycle length of about $50$ km.  First, perturbation theory should handle the perturbation through a cycle as that implies a single main interaction of the acoustic and internal waves.  As the perturbation has a very complicated, seemingly random, appearance, it may generate more or less random matrix element phases as desired in (ii) above.   Second, the next cycle should lead to a roughly independent interaction as the internal waves separated by $50$ km are for the most part independent.  That would satisfy (i) above.  Finally, all parts of the wave evolution have interacted with the internal waves near the surface.  This would seem to be essential (necessary, though not guaranteed to be sufficient) for reducing dynamical correlations [(iii) above].  In fact, as shown in Refs.~~\onlinecite{Hegewisch12,Hegewisch10}, this is the least well satisfied criterion.  We shall neglect residual dynamical correlations to begin with, but realize that if this approximation is not good, we will find statistical deviations ahead.  If however, none show up, then that indicates that the information contained in these correlations is effectively being lost in the long range propagation. In addition to that knowledge, it is not necessary to return to the ensemble construction to attempt to incorporate them.

By propagating the parabolic equation or by evaluating Eq.~(\ref{av}), it is found that the off-diagonal matrix elements $A_{mn}$ $(m \ne n)$ do, in fact, behave as though their phases are random by about $50$ km of propagation.  They also behave like complex Gaussian random numbers.  Therefore, only their variance needs to be specified (the mean of Eq.~(\ref{av}) is zero),
\begin{eqnarray}
\label{eq:A}
A_{mn}(k) =\sigma_{mn}(k) z_{mn}(k)  \ ,
\end{eqnarray}
where  $z_{mn}(k)$ is a complex Gaussian random variable of zero mean and unit variance (with $\langle z_{mn}(k) z_{m^\prime n^\prime}(k)\rangle = \delta_{mm^\prime} \delta_{nn^\prime}$), and the $k$-dependence has been emphasized because it must be accounted for in the construction of a  timefront.  For $m=n$, the Gaussian random variable, $z$, is real with zero mean and unit variance, but otherwise Eq.~(\ref{eq:A}) applies.  
The variance can be calculated analytically using Eqs.~(\ref{vimn}), (\ref{av}) and the expressions in Appendix~\ref{ap:soundspeed}, which give
\begin{eqnarray}
\epsilon^2\sigma^2_{mn}(k) &=&
\frac{k^2r_b^2}{16} \sum_l \left[ {\rm sinc}^2\left(\frac{\omega^{+,l}_{mn}r_b}{2}\right)+ {\rm sinc}^2\left(\frac{\omega^{-,l}_{mn}r_b}{2}\right)\right] \nonumber \\
&&\times \sum_j  \left|V_{mn}^{j,l}(k)\right|^2  \ , 
\end{eqnarray}
where
\begin{eqnarray}
\label{defs}
\omega^{\pm,l}_{mn} &=& k(E_m-E_n) \pm k_l \ , \nonumber \\
V^{j,l}_{mn}(k) &=& \int  {\rm d}z V_j(z;k_l) \psi^*_m(z;k) \psi_n(z;k)  \ ,
\end{eqnarray}
and ${\rm sinc}(x) = \sin x/x$.  The sinc-functions can be thought of as smeared $\delta$-functions that pick out the most important horizontal wave vector  components for that particular matrix element, which is crudely a function of $|m-n|$.  The $ \sum_j \left|V_{mn}^{j,l}(k)\right|^2$ determines the weighting and is calculated by numerically solving for the modal eigenfunctions.  For some potentials, such as the harmonic oscillator, the elements could be calculated analytically.  This expression for the variance generates building-block, unitary propagation matrices that have matrix element statistics which agree very well with those from solutions of the parabolic equation~\cite{Hegewisch12}.   

The integrand of Eq.~(\ref{vimn}) has three oscillatory components.  In order for the integral not to mostly self-cancel, there must be an oscillatory component of $V_1(z,r)$ that matches to some extent differences between the oscillations of $\psi_m(z;k)$ and $\psi_n(z;k)$.  The higher frequency components of $V_1(z,r)$ fall off in intensity, and therefore one anticipates that as the difference $|m-n|$ grows, the matrix elements must decay in strength.  This was found in Ref.~~\onlinecite{Hegewisch12} to result in a mostly power-law decay as a function of $|m-n|$ and to be roughly independent of the mean index $|m+n|/2$.

Perturbatively, the diagonal element $A_{mm}(k)$ is mostly responsible for introducing a dephasing  away from the phase $\exp(-ikE_mr)$ that arises from the eigenmodes of the unperturbed system.  In principle, it has both a small drift (from higher order terms in $\epsilon$) and a fluctuating part.  Here, we incorporate only a fluctuating part in the random matrix ensemble.

\section{Acoustic timefronts}
\label{results}

In order to carry out the ensemble construction of timefronts according to Eqs.~(\ref{tf}), (\ref{eq:product}), there is one final correlation to consider.  The value of $A_{mn}(k)$ deduced from a dynamical system, say using Eq.~(\ref{av}), at two different values of the wave vector $k$ could be correlated.  In fact, at a minimum there must be very strong correlations for two values of $k$ which closely approach each other.  This alone implies that the values of $z_{mn}(k_1)$ and $z_{mn}(k_2)$ for a single member of the ensemble (corresponding vaguely to a single realization of the internal wave field) should be taken as being nearly identical for very small $|k_1-k_2|$ with some fall off as the difference grows (not accounting for further types of correlations).  A priori, the fall off could also depend on $|m-n|$ as well.  Numerically, there is considerable decay of correlation over the range of wave vectors represented in a timefront and there did not appear to be a simple rule to describe its behavior.  Nevertheless, it turns out that the gross approximation of taking $z_{mn}(k)$ identical for all values of the wave vector is a far better starting point than the opposite ``white noise'' limit in which each $z_{mn}(k)$ is taken as completely statistically independent.  The results shown below use the former, perfectly correlated choice.  Again, if statistical deviations are found that can be tracked back to this crude approximation, then that information can be deduced from the propagation.  On the other hand, if one does not see statistical deviations, then the information about the correlation decay as a function of $|k_1-k_2|$ is being lost by the propagation.

\subsection{Sample timefronts}

We compare the appearance of a single timefront generated using the random matrix ensemble technique with another generated using the parabolic equation.  The question is whether the two timefronts would be similar in their overall structure, but not in the details of their fluctuations.  The former (or ensemble calculation) proceeds by applying Eqs.~(\ref{tf}), (\ref{eq:product}), (\ref{eq:unitary}), (\ref{eq:A}-\ref{defs}) following the prescriptions and choices described in the text.  The latter (or parabolic equation) calculation relies on Eqs.~(\ref{eq:parabolic}), (\ref{eq:V}-\ref{eq:colosibrown_c}), (\ref{eq:timefront}) for a single realization of the internal wave field.

\begin{figure}
\centering
     \subfigure{
          \label{fig:pt_sample_timefront_em}
          \includegraphics[width=3.5in,angle=0]{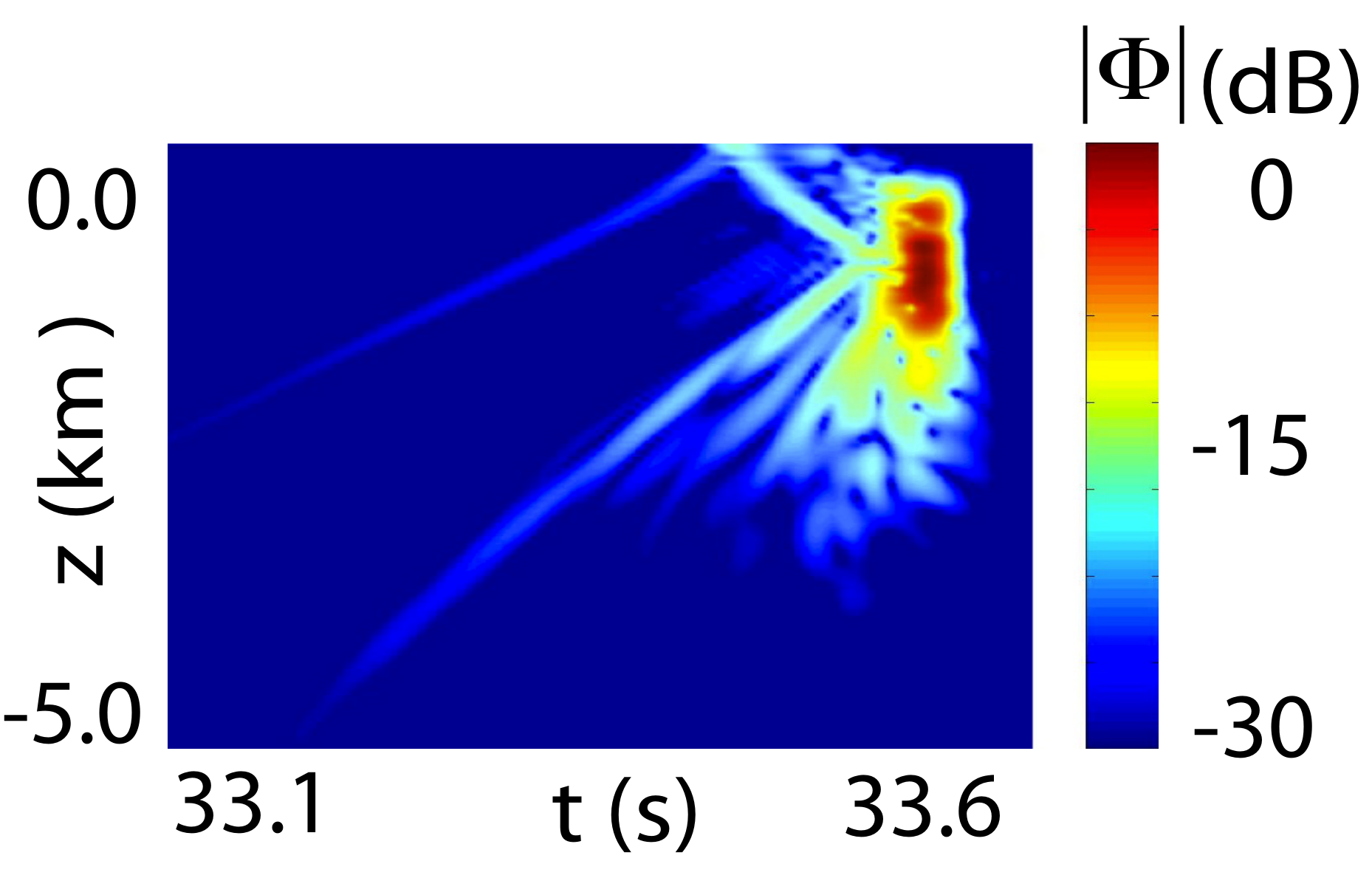}}\\
     \subfigure{
          \label{fig:pt_sample_timefront_prop}
          \includegraphics[width=3.5in,angle=0]{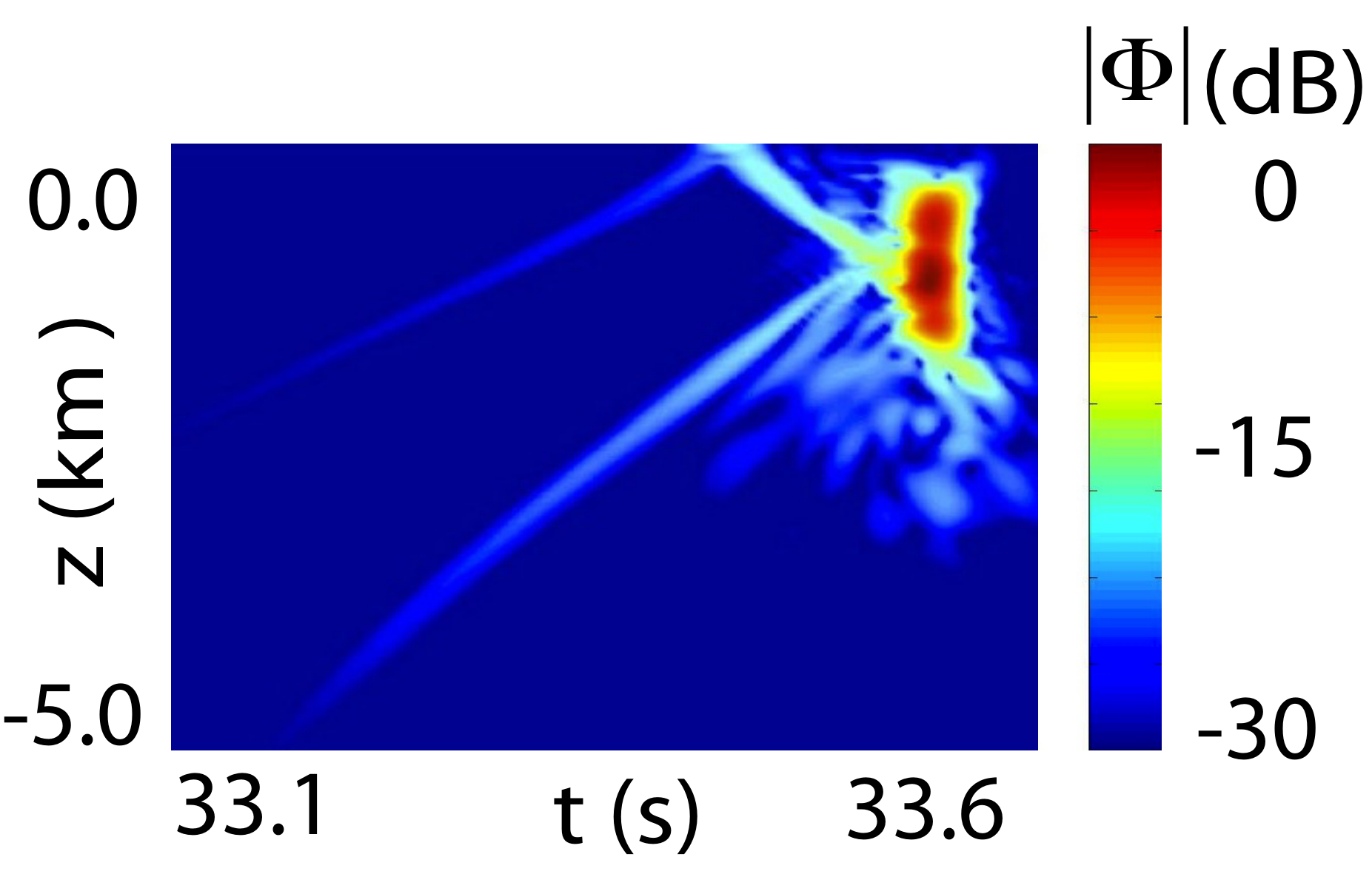}}
     \caption[Comparison of sample timefronts for $50$ km with $mu_{n,n} = 0$]{
Comparison of sample timefronts from a random matrix ensemble construction and parabolic equation propagation for $50$ km.  The magnitudes $|\Phi|$ of the timefront are plotted in decibels with respect to the largest value of $|\Phi|$ as a color density plot in depth $z$ and time $t$.
The upper plot is a sample $|\Phi|$ generated from a single member of the ensemble model.
The lower plot is a sample $|\Phi|$ from propagation through a single internal wave field.}
\label{fig:timefront_overlap_ptstats_50}
\end{figure}

\begin{figure}
\centering
     \subfigure{
          \label{fig:timefront_em_250}
          \includegraphics[width=3.5in,angle=0]{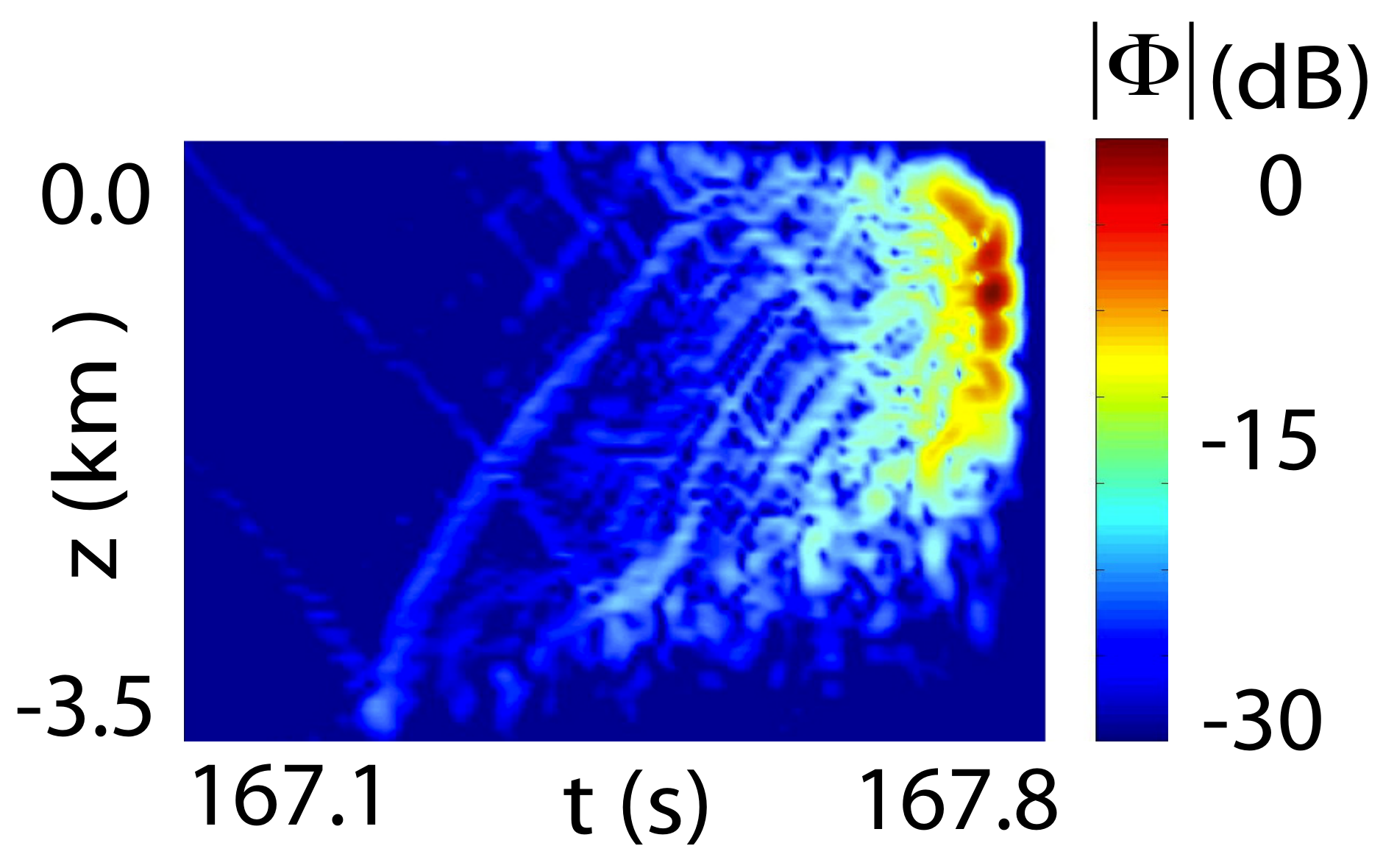}}\\
     \subfigure{
          \label{fig:timefront_prop_250}
          \includegraphics[width=3.5in,angle=0]{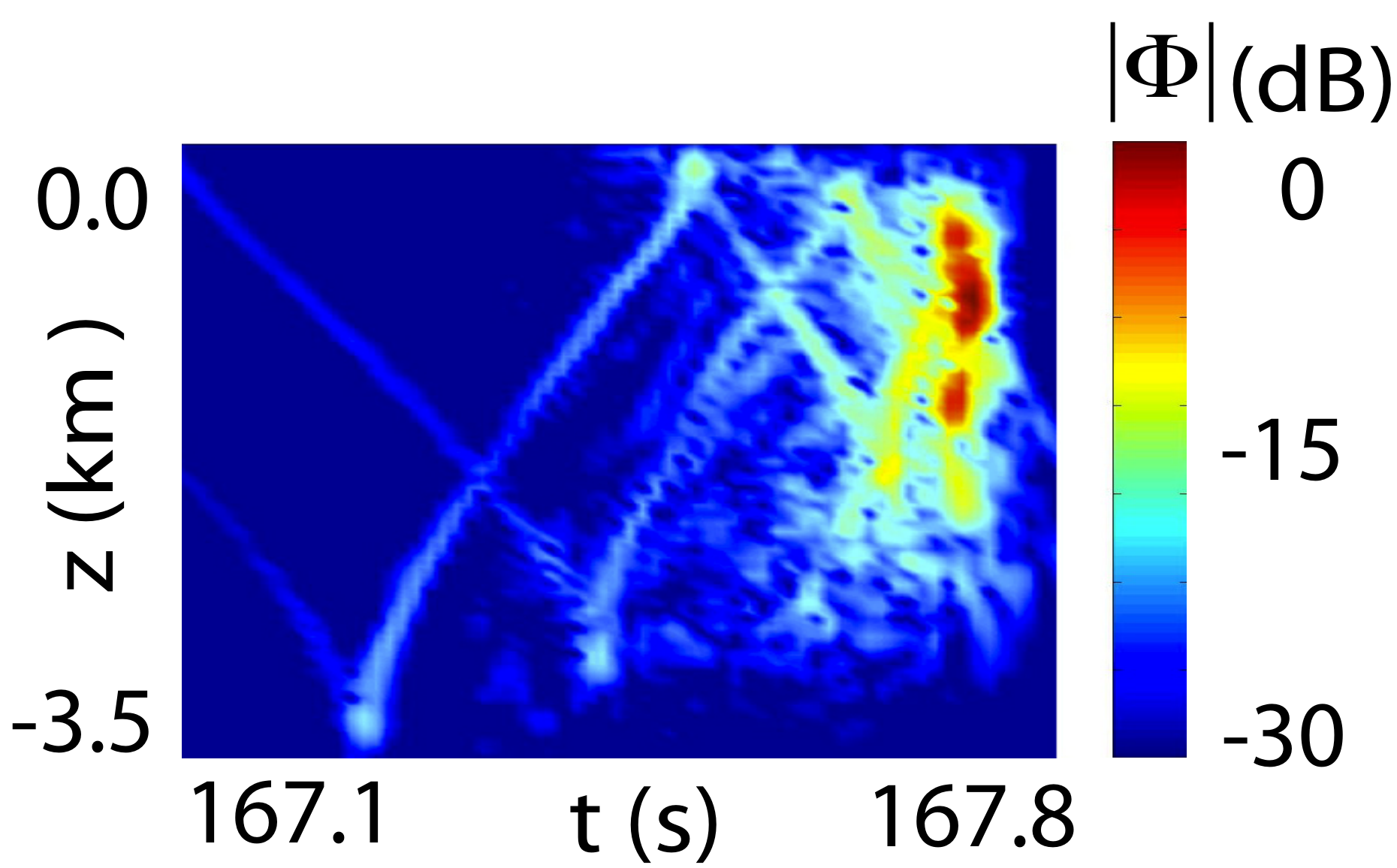}}
     \caption[Comparison of timefronts for $250$ km]{
Comparison of sample timefronts from a random matrix ensemble construction and parabolic equation propagation for $250$ km.  The magnitudes $|\Phi|$ of the timefront are
plotted in decibels with respect to the largest value of $|\Phi|$ as a ca color density plot in depth $z$ and time $t$.  The upper plot is a sample $|\Phi|$ generated from a single member of the ensemble model.  The lower plot is a sample $|\Phi|$  from propagation through a single internal wave field is shown.}
\label{fig:timefront_overlap_ptstats_250}
\end{figure}

For the propagation distance of a single building block, see the comparison given in Fig.~\ref{fig:timefront_overlap_ptstats_50}.  Both realizations span roughly the same decibel range 
and possess similar global structures.  The branches cover the same locations with about the correct intensities.  The variations of the acoustic power with depth in the finale are quite similar.  The fine structures that contain the fluctuations are similar in character, but differ in the detailed values.  The next test of the random matrix ensemble, in particular the building block structure, is to compare timefronts for propagation to longer ranges, which tests Eq.~(\ref{eq:product}); see Fig.~\ref{fig:timefront_overlap_ptstats_250}.  For this particular case, 
globally, the structure is again similar, however the fluctuations appear to be somewhat stronger in the ensemble construction.  Whether or not this is a part of the statistical variation to be expected or an excess in the ensemble construction requires evaluating ensemble statistical measures.

\subsection{Average intensity timefronts}

\begin{figure}
\centering
     \subfigure{
          \label{fig:Iavg_iw_50}
          \includegraphics[width=3.5in,angle=0]{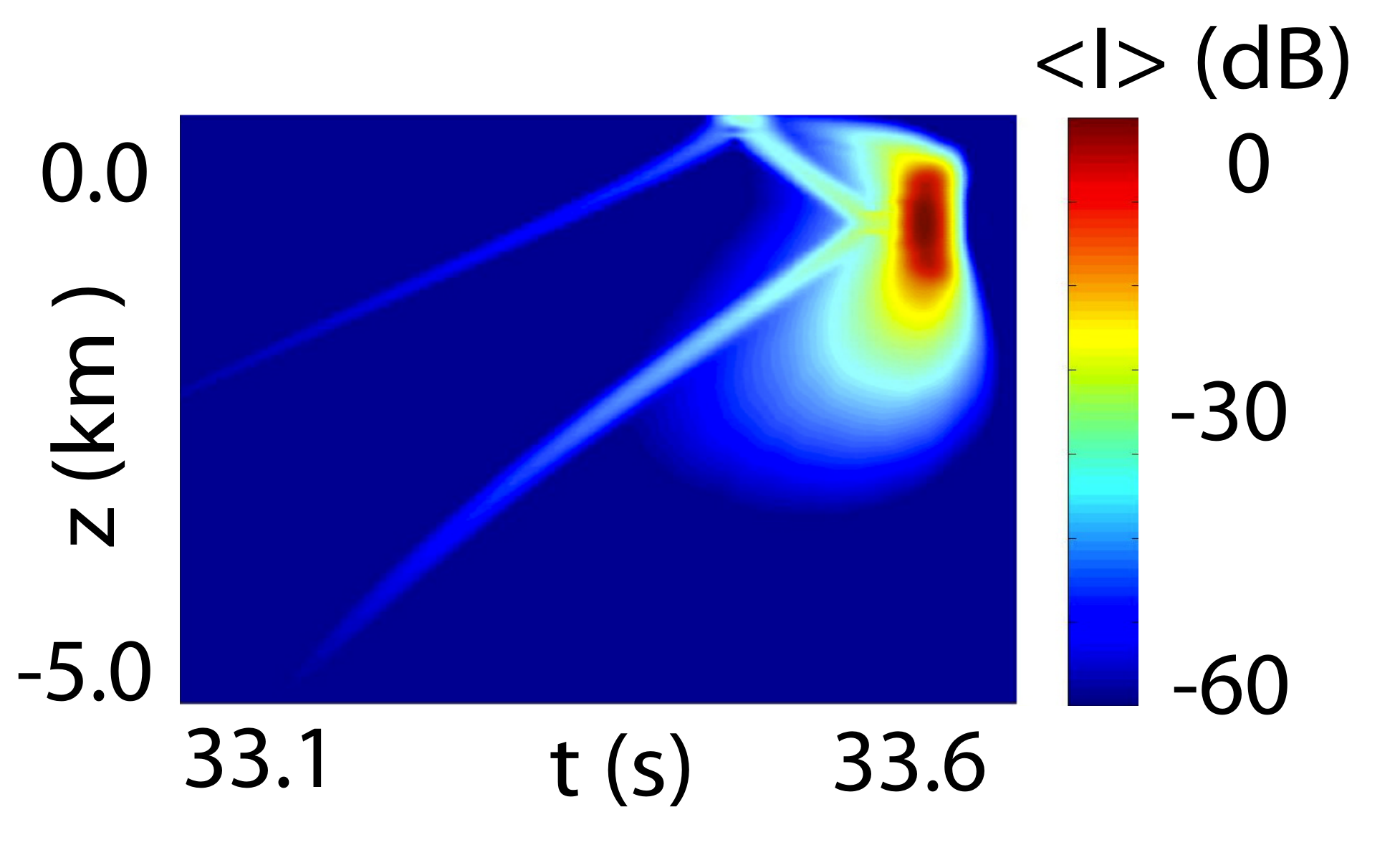}}\\      
\subfigure{
          \label{fig:Iavg_ensemble_50}
          \includegraphics[width=3.5in,angle=0]{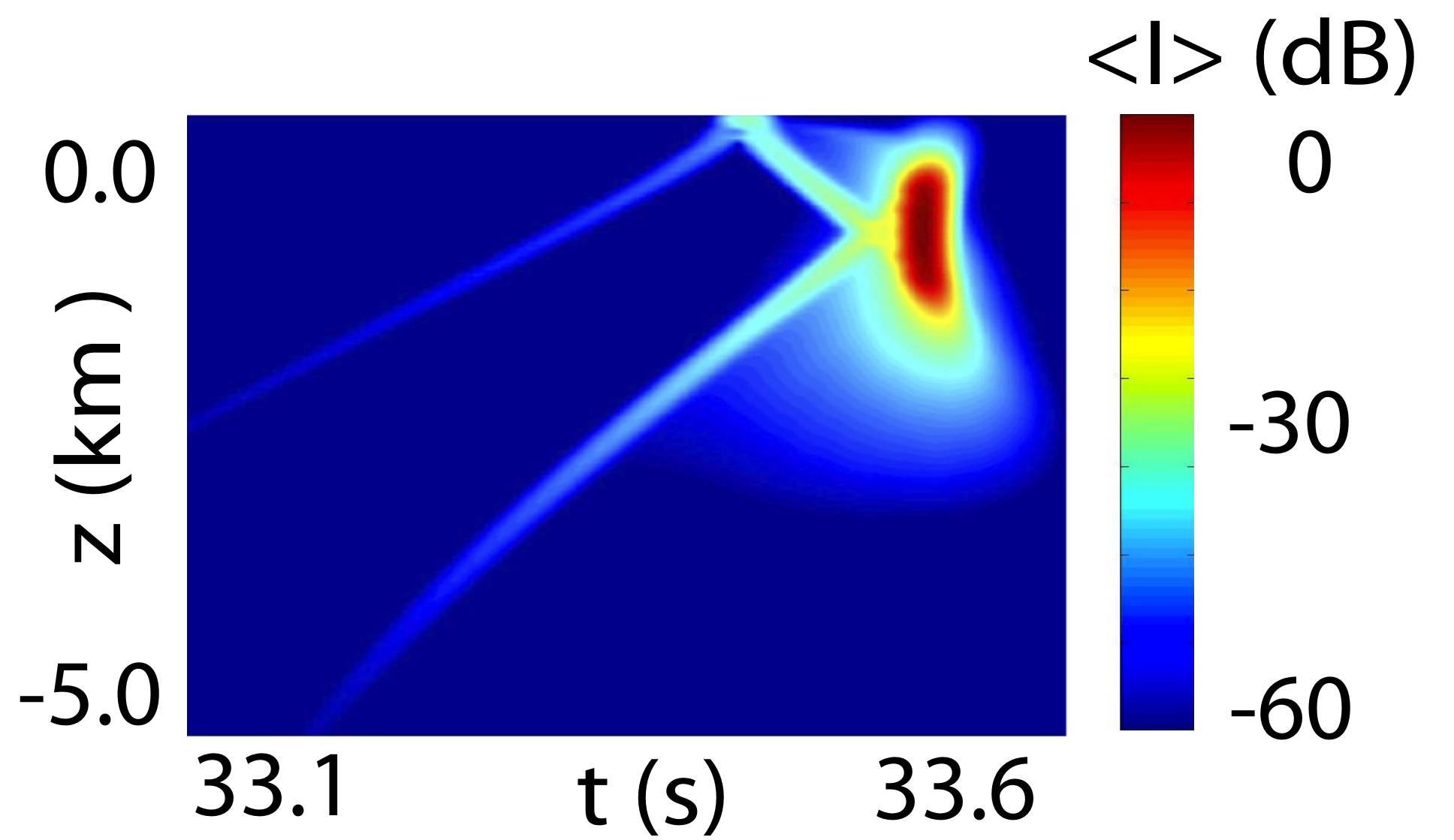}}
     \caption[Comparison of average intensity timefront for $50$ km]{
The average intensity timefront $\langle I\rangle= \langle |\Phi|^2\rangle$ is shown as a color density plot with depth $z$ and time $t$.  (Upper) The average is taken over timefronts resulting from ensemble calculations utilizing $1000$ independent random matrices for $50$ km. (Lower) The average is taken over timefronts from the propagation through $2500$ independent sound speed models to $50$ km.}
     \label{fig:Iavg_50}
\end{figure}

Consider the average intensity~\cite{Flatte83} $\langle I (z,r;t)\rangle$ where the brackets $\langle \dots \rangle$ indicate the ensemble average for the random matrix ensemble and the average over the internal wave realizations for the parabolic equation propagation,  
\begin{eqnarray}\label{eq:avgI}
\langle I(z,r;t) \rangle = \frac{1}{N}\sum^N_{l=1} |\Phi_l(z,r;t)|^2 \ .
\end{eqnarray}
Here $N$ is the number of internal wave realizations or random matrix ensemble members and is chosen large enough to have small statistical error ($N$ between $1000$ and $2500$ in our calculations); the subscript $l$ indicates which realization.  The average intensity timefronts from the random matrix ensemble and parabolic equation propagations for $50$ km are compared in Fig.~\ref{fig:Iavg_50}.  In these figures, one sees 
the two previously mentioned global structures, the branches and the complicated finale.  The branches are almost perfectly similar to their appearance in a timefront unperturbed by internal waves, other than perhaps a small drift in time.  The finale is a region of overlapping branches, so much so that there is little trace of their origins.  The average intensities of the random matrix ensemble and parabolic equation are quite similar, although there appears to be a few small differences in the finale, such as extra mixing might create, i.e. more acoustic power at lower depths.   Otherwise, there is enough similarity between the two constructions that the random matrix ensemble could be used to replace the parabolic equation as far as the average intensity is concerned.

\begin{figure}
\includegraphics[width=3.3 in,angle=0]{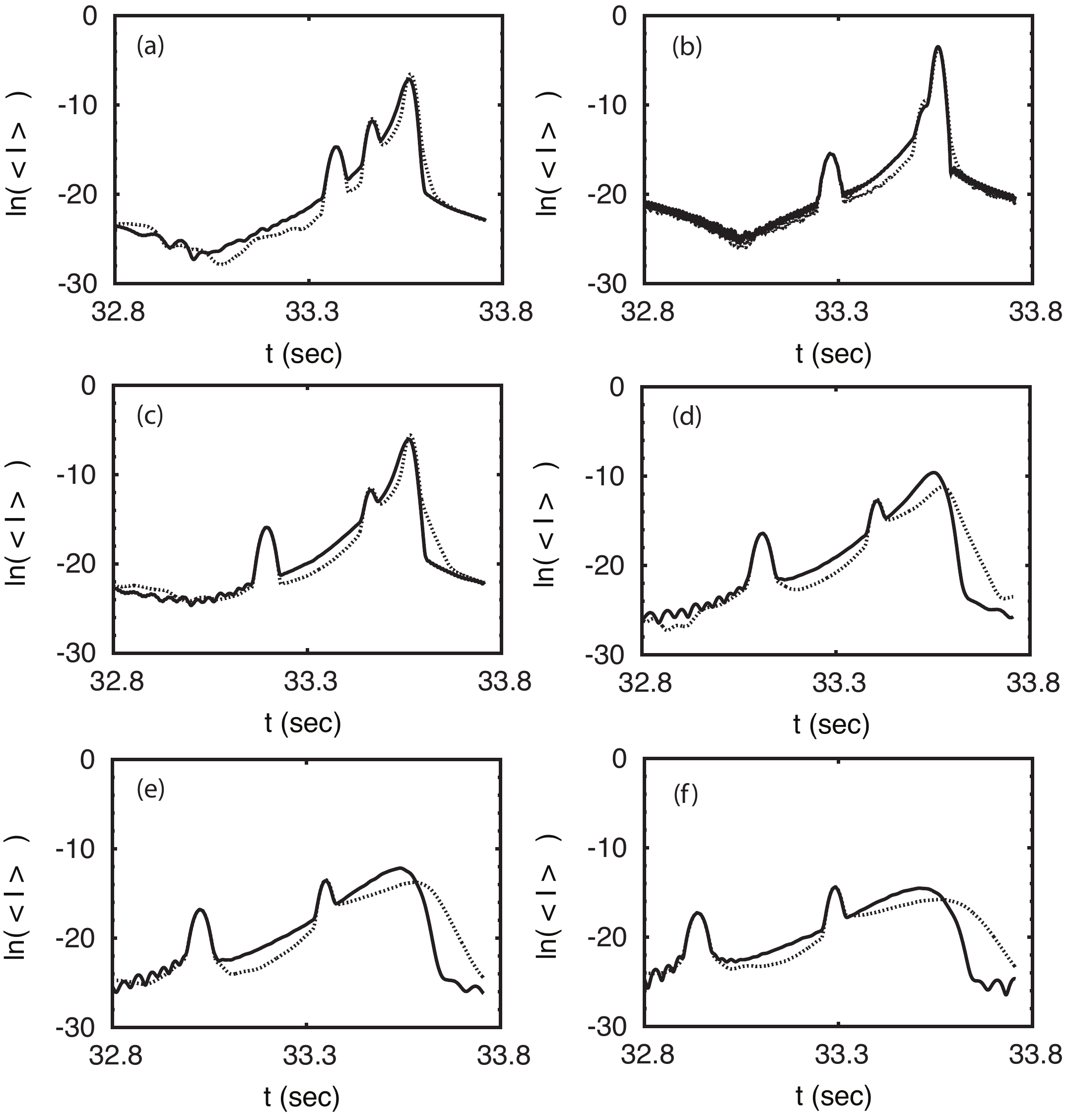}  
\caption[Comparison of traces of average intensity timefronts for $50$ km]{
Traces of the average intensity are shown for several final depths $z$ (0.5, 1.0, 1.5, 2.0, 2.5, 3.0 km, respectively) as a function of time $t$.  The average of the ensemble model timefronts (solid lines) is taken  from $1000$ independent random matrix ensemble members for $50$ km.  The average of the propagated timefronts (dotted lines) is taken from the propagation through $2500$ independent sound speed models to $50$ km.}
\label{fig:Iavg_traces_50}
\end{figure}

A finer look at the comparison is shown in Fig.~\ref{fig:Iavg_traces_50} by fixing the depth and range, and plotting 
the resultant time traces as a function of time (i.e.~section through the timefront).  The ensemble captures the locations and magnitudes of the branches exceedingly well.  In the gaps between the branches, there is a background determined by the internal wave scattering, i.e.~mode-mixing.  It peaks near $(z,t) = (-1 , r/c_0)$ and decays in both depth and time from this peak.  The rate of decay in both depth and time of this structure gives a signature for the multiple scattering of the internal waves.  It appears to be weakly enhanced in the random matrix ensemble results relative to the parabolic equation propagation results. The enhancement appears to increase slowly with increasing depth of the trace and possibly with decreasing time.  As a result, the earlier branches are not quite as well resolved.  The origin of this weak distinction between the random matrix timefront construction and the parabolic equation one has not yet been identified.  One possible explanation of this weak discrepancy is the approximation of perfect coherence as the value of the wave vector $k$ changes ($z_{mn}(k_1)=z_{mn}(k_2)$ for all $k_1,k_2$).  This would exaggerate the effect of the perturbation.  Less likely, but also a possible explanation, is ignoring the weak dynamical correlations that must be present in any continuous dynamical system.  Identifying the precise cause of the small differences and improving the ensemble timefront construction is beyond the scope of this paper and is left for future work.  Regardless of the origin, it may be possible to make a small rescaling effectively reducing the $\{\sigma^2_{mn}(k)\}$ from their perturbation theory values and maintain just as simple an ensemble as the current one.

\begin{figure}
\centering
     \subfigure{
          \label{fig:Iavg_iw_1000}
          \includegraphics[width=3.5in,angle=0]{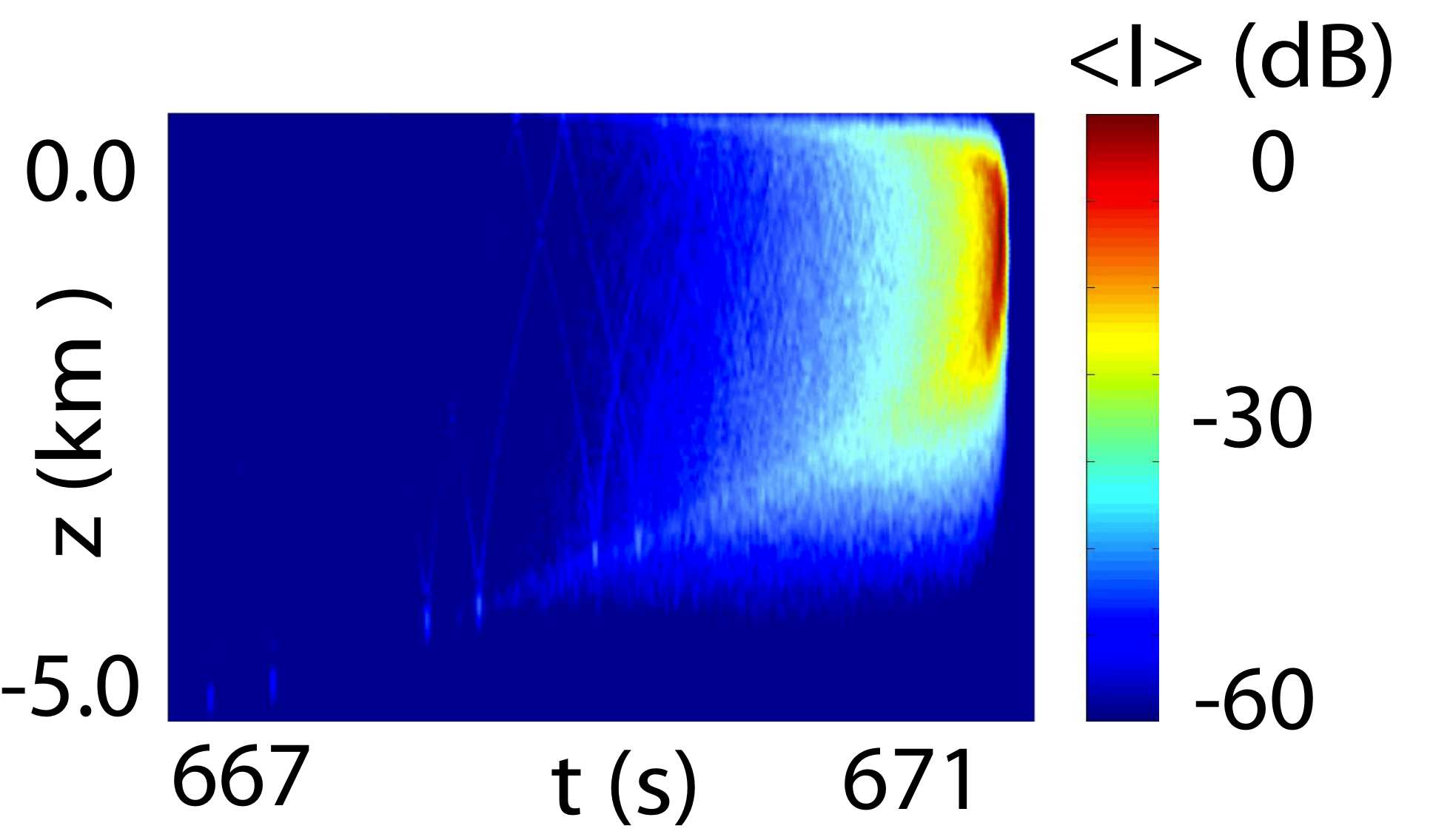}}\\
     \subfigure{
          \label{fig:Iavg_ensemble_1000}
         \includegraphics[width=3.5in,angle=0]{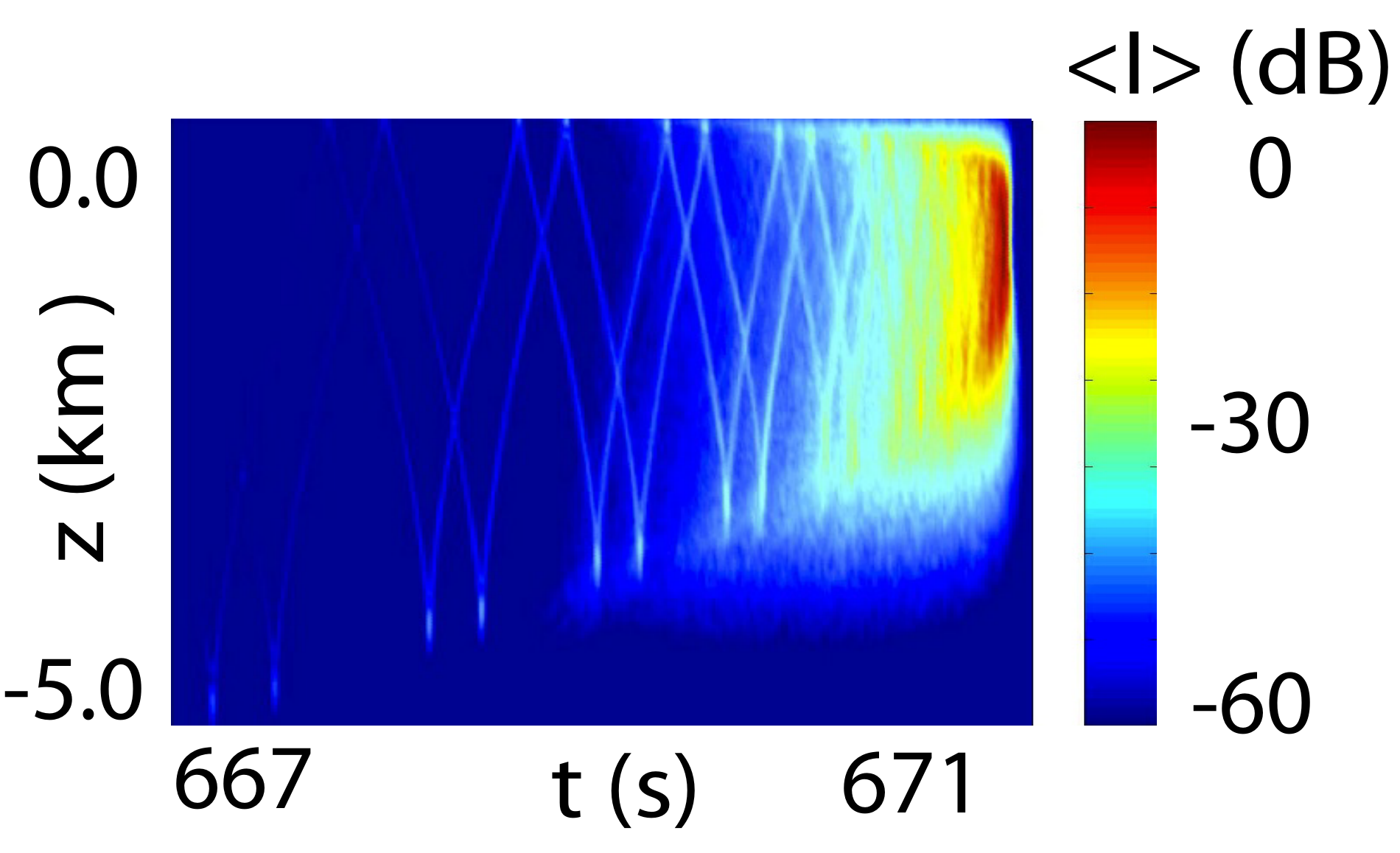}}
     \caption[Comparions of average intensity timefronts for $1000$ km]{
The average intensity timefront $\langle I\rangle= \langle |\Phi|^2\rangle$ is shown as a color density plot with depth $z$ and time $t$.  (Upper) The average intensity timefront results from $25$ independent random matrix ensemble members propagated for $1000$ km.  (Lower) The average intensity timefront is from the propagation through $28$ independent sound speed realizations to $1000$ km.}
     \label{fig:Iavg_1000}
\end{figure}

\begin{figure}
\includegraphics[width=3.5 in]{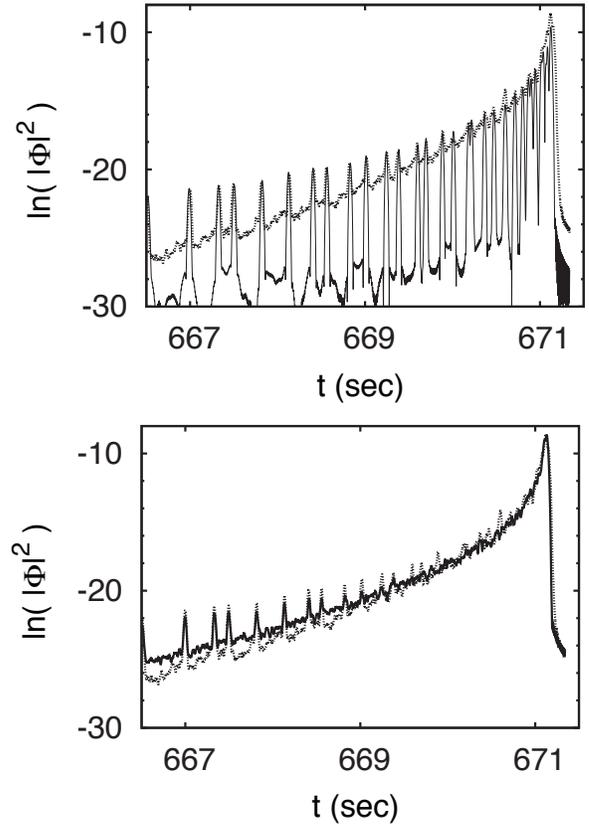}
\caption[Comparison of traces of average intensity timefronts for $1000$ km]{
Traces of the average intensity $\langle I\rangle= \langle |\phi|^2\rangle$ are shown for a single final depth $z=1.5$ km as a function of time $t$.  (Upper) A comparison of the average propagated timefront traces with and without internal waves (solid line without internal waves).  (Lower) The average of the ensemble timefronts (solid line) is from $1000$ independent members for $1000$ km.  The average of the propagated timefronts (dotted line) is from $2500$ independent internal wave realizations propagated to $1000$ km.}
\label{fig:Iavg_traces_1000}
\end{figure}

To test how the multiplicative structure of the ensemble fares for the average intensity, one must again look at longer ranges.  In Fig.~\ref{fig:Iavg_1000} an average timefront comparison is shown for propagation to $1000$ km.  
The average intensity timefronts from the ensemble model capture the finale region very well both in decay with time and with depth, and the positions of the branches are excellent.  However, the excess scattering results in the branches being less visible in the color density plot.  Looking at the time traces  in Fig.~\ref{fig:Iavg_traces_1000} gives a sharper view just as it did for Fig.~\ref{fig:Iavg_traces_50}.  In the upper plot, the internal waves are turned off to illustrate that in between the branches, the signal is many orders of magnitude decreased and one effect of internal waves is to infill a structure between the branches.  In the lower plot, the random matrix ensemble captures the fill-in very well, but shows some excess intensity
in the gaps relative to the parabolic equation and the magnitudes of the branches are reduced somewhat though still in the correct positions.  We suspect that improving the building block ensemble would alleviate this slight discrepancy at longer ranges and it is not due to the the complete independence assumption of adjacent building blocks.

\subsection{The mixing front}

In the ensemble averaged intensity, $\langle I(z,r;t) \rangle$, the branches are largely unchanged from the unperturbed system.  That suggests looking at the ensemble averaged intensity change
\begin{equation}
\label{eq:ens_pred}
\delta I_{\epsilon}(z,r;t) =\langle I_\epsilon (z,r;t) \rangle -  I_{\epsilon=0}(z,r;t)   \ ,
\end{equation}
whose functional dependence reflects the scattering due to the internal waves.  This is related to the mode-mixing, albeit over a range of wave vectors, because the early arrivals represent wave energy in the higher modes (steeper propagation - reverse dispersion case).  If there were complete mixing the branches would disappear, and the early and late parts of the timefronts would be statistically and structurally indistinguishable.  As this is not the case, it is possible to deduce information about the mode-mixing from $\delta I_{\epsilon}(z,r;t)$.  We call this the mixing front, whose decay toward the early part of the arrival (as a function of $r/c_0 - t$) reflects the modal decomposition of the initial signal convolved with the mixing from the scattering.   For short enough propagation ranges, this should be dominated by the modal decomposition of the initial signal, but as the range increases the power-law decay of the unitary matrix elements as a function of $|m-n|$ must eventually dominate.  More generally, the average intensity of a large number of well-separated-in-time, long-range, ocean acoustic measurements would allow one to deduce information about the sum of all the scattering processes relevant to those measurements, i.e.~that is if the experimental noise levels are sufficiently controlled that the mixing front can be made to emerge from the average.

\begin{figure}
\includegraphics[width=3.5 in]{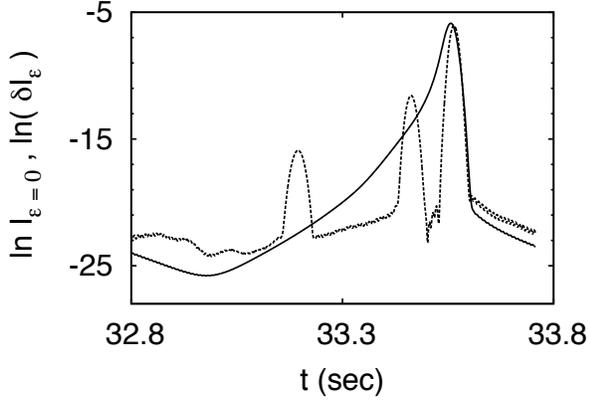}
\caption{The solid line depicts the mixing front $\delta I_{\epsilon}(z,r;t)$.  The peaks have disappeared.  The average intensity without internal waves shows where the peaks  of the full timefront would be.}
\label{fig:prediction_time}
\end{figure}

The random matrix ensemble for the acoustic propagation can be used to derive an expression for $\delta I_{\epsilon}(z,r;t)$ without having to perform full propagations over a large ensemble of internal wave fields.  Starting by approximating the unitary propagation matrix elements as 
\begin{eqnarray}
U_{m,n}(r_b;k) \approx e^{-ikr_b E_m}\left( \delta_{m,n} - 2i \sigma_{A_{m,n} }  z_{m,n} \right)
\end{eqnarray}
and performing the ensemble averages using the statistical properties of $z_{m,n}$, the ensemble averaged intensity change, i.e.~the 'mixing front', takes the form
\begin{eqnarray}
\delta I_{\epsilon} =&&
\frac{\epsilon^2}{2 \pi \sigma_k^2 r_b} \sum_{m,n} \left| \int_{-\infty}^{+\infty} dk ~  e^{ - i k c_0\left(  t - \frac{r_b}{c_0} \right)}
\exp{\left[\frac{-(k-k_0)^2 }{2 \sigma_k^2}\right]} \right.\nonumber \\
&& \left. 2 \sigma_{A_{m,n}}(k) e^{-ikr_b E_m } a_n(k) \psi_m(z;k) \right|^2 \ . \nonumber \\
\label{eq:Ieps}
\end{eqnarray}
The mixing front for a single depth is illustrated in Fig.~\ref{fig:prediction_time}.
The effect of the mixing front on the timefront is to hide most of the structure of the unperturbed arrivals in the finale region, with only a slight broadening of the unperturbed early arrivals.  This creates the appearance of a smearing of the arrivals in the finale region.  Thus, the random matrix ensemble model gives an explanation of the appearance of the arrivals smearing in the finale region and why the unperturbed arrivals are still visible in the early arrivals of the timefront.   As a practical consideration, experimental determination of the mixing front would be enhanced by a better signal-to-noise for shorter range propagations.  However, longer range propagation would be more dominated by the power-law behaviors and the mixing front would emerge more clearly.  A balance of these competing factors would determine the best ranges for testing mixing front predictions with measurements.

\begin{figure}
\includegraphics[width=3.5 in]{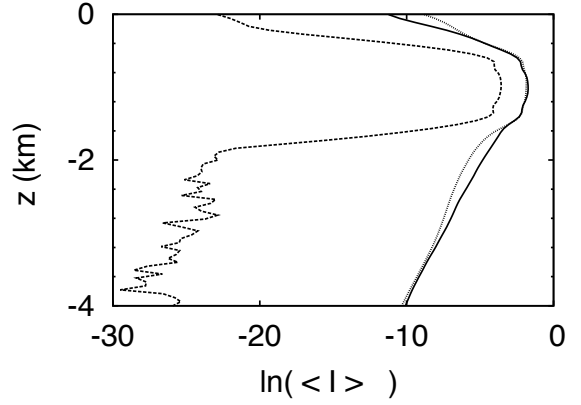}
\caption{The average intensity $\langle I \rangle$ of the acoustic timefront 
is plotted with depth $z$ at the 
time $t=r/c_0$ for the ensemble model prediction $\langle I_\epsilon \rangle$ from Eq.~(\ref{eq:ens_pred}) (Solid line), for a statistical average of intensities from 
several perturbed propagations (Dotted line) and 
for the unperturbed propagation $I_{\epsilon=0}$ (Dashed line). }
\label{fig:prediction_depth}
\end{figure}

Figure~\ref{fig:prediction_depth} illustrates the depth dependence of the mixing front 
of Eq.~(\ref{eq:ens_pred}).  As seen in the figure, $\delta I_{\epsilon}$ results in a greater intensity for the arrivals in the finale region (compared to the unperturbed propagation) and a slower decay in depth.  The decay in depth may transition from an exponential to a power-law decay as internal waves are taken into account.  This prediction closely approximates the average intensity computed from perturbed propagations.  Thus the random matrix ensemble also gives an explanation of the increase in vertical extent of the arrivals of finales of the perturbed timefronts over those of the unperturbed timefronts.

\section{CONCLUSIONS}
The random matrix ensemble for the unitary propagation matrices capture the statistical properties of the propagation~\cite{Hegewisch12} quite well without fitting parameters.  The ensemble is derived using first order wave perturbation theory.   The ensemble predicts the presence of a mixing front in the acoustic timefronts.  This front is consistent with the long-known presence of a smeared region of arrivals in the finale region of the timefronts and the persistence of branches in the early timefront. 
Its characteristic features are the decay with depth and with time from a peak near the horizontal arrival. 
These features seen in the ensemble agree with parabolic equation propagation. 

Though perturbation theory alone would only be able to make predictions about the unitary propagation matrices to very short ranges, the statistical properties predicted using perturbation theory seem to be more robust over the course of the propagation to much longer ranges (i.e. 3000 km).  The successful use of first order perturbation theory in capturing useful statistics for the long range propagation suggest that perturbation theory may have statistical use for other fields as well. 

The realization that internal wave effects cause a mixing front in the acoustic timefront of arrivals may lead to a new way of deducing information about the internal state of the ocean. This front has 
characteristic features in depth and time that depend on the properties of the mode mixing. 
Their parameters are controlled by the strength and structure of the internal wave sound speed perturbations.  For example, measurements of the vertical extent of arrivals in the finale region have already been analyzed~\cite{uffelen09,uffelen10}.  Comparing to the predicted properties of a mixing front may give greater understanding of the ocean environment.

There are many possible directions for future research.  More careful identification of information lost and retained and what effects remain would be very interesting.  There is also the possibility of extending the construction of random matrix ensembles to broader classes of ocean acoustics systems where absorption plays a role.  More work should also be done with the fluctuation properties of the timefronts and attempts made to have analytic random matrix predictions for their behaviors.

\begin{acknowledgments}
We wish to acknowledge the support of the National Science Foundation through the grant PHY-0855337.  The computational resources for this research were supported in part by the National Science Foundation through TeraGrid resources provided by NCSA. TeraGrid systems are hosted by Indiana University, LONI, NCAR, NCSA, NICS, ORNL, PSC, Purdue University, SDSC, TACC and UC/ANL.
\end{acknowledgments}

\appendix

\section{Implementation of internal wave model}
\label{ap:soundspeed}

The efficient numerical scheme devised by Colosi and Brown~\cite{Colosi98} generates a statistical ensemble of internal wave effects for the sound speed model through the equation:
\begin{eqnarray}
\label{a1}
&&\epsilon V_1(z,r) =
\frac{24.5}{g}\frac{2B}{\pi} N_0^2 \sqrt{\frac{E \Delta k_l}{M}}
\exp(-3z/2B) \hspace{0.1in} \\
&&\times\sum_{j=1}^{j_{max}}
\sum_{l=1}^{l_{max}} \sin(j\pi\xi(z))
\sqrt{\frac{I_{j,k_l}}{j^2+j_*^2}} \cos\left( \phi_{jl}+k_l r\right) \ , \nonumber
\end{eqnarray}
where the depth dependence and weighting of the $j^{th}$ internal wave mode is
\begin{eqnarray}
\label{a2}
 V_j(z;k_l) &=& \frac{24.5}{g}\frac{2B}{\pi} N_0^2 \sqrt{\frac{E \Delta k_l}{M}} \exp(-3z/2B) \times \nonumber\\
 && \sin(j\pi\xi(z)) \sqrt{\frac{I_{j,k_l}}{j^2+j_*^2}} \nonumber \\
k_j I_{j,k_l} &=& \frac{1}{\beta^2 +1}+
\frac{1}{2}\frac{\beta^2}{(\beta^2
+1)^{\frac{3}{2}}} \ln\left(\frac{\sqrt{\beta^2+1}+1}{\sqrt{\beta^2+1}-1} \right) \ , 
 \nonumber \\
\end{eqnarray}
\noindent and $\xi(z) = e^{-z/B}-e^{-H/B}$. 
A single random seed generates a sequence for the random phases, $\phi_{jl}\in[0,2\pi)$,  of each internal wave with vertical mode, $j$, and horizontal wavenumber, $k_l$.  The ensemble has the statistical properties of the Garrett-Munk spectrum~\cite{Garrett79}. The full Garrett-Munk energy of
$E= 6.3 \mbox{ x }  10^{-5}$ has been used in all calculations.  The calculations are done for a latitude of $30^\circ$ so that the inertial frequency is $f_i=1$ cycle per day.  The buoyancy profile is assumed to have the form $N(z)=N_0 e^{-z/B}$, where $N_0=$ 1 cycle per 10 min is the buoyancy frequency at the surface.  The depth of the ocean is considered to be $H=5.0$ km, although the propagation depths have been extended to the vertical region $[-3,10]$ km~\cite{Hegewisch05} for our calculations.

The particular functional forms and constants used in this paper are as used by Colosi and Brown.  
Some of these forms and constants have already been identified in the body of the paper (i.e. near Eq.~(\ref{eq:colosibrown_c})), while the others are listed here:  gravitational constant $g=9.81$ m/s$^2$, mode scaling number $M=(\pi j_* -1)/2 j_*^2$ and principle mode number $j_* = 3$. The horizontal internal wave number $k_l$ are taken to be $512$ values equally spaced by $\Delta k_l$ for
$k_l\in 2\pi[0.01,1.0]$ radians per km.  In the expression for $I_{jl}$, $k_j = f_i\pi j/N_0 B$ and the ratio $\beta =k_l/k_j$.

\bibliography{oceanacoustics,quantumchaos,nano,rmtmodify}

\begin{thebibliography}{10}
\newcommand{\enquote}[1]{``#1''}
\expandafter\ifx\csname url\endcsname\relax
  \def\url#1{\texttt{#1}}\fi
\expandafter\ifx\csname urlprefix\endcsname\relax\def\urlprefix{URL }\fi
\providecommand{\bibinfo}[2]{#2}
\providecommand{\noopsort}[1]{}
\providecommand{\switchargs}[2]{#2#1}

\bibitem{Flatte79}
\bibinfo{author}{S.~M. Flatt\'e}, \bibinfo{author}{R.~Dashen},
  \bibinfo{author}{W.~H. Munk}, and \bibinfo{author}{F.~Zachariasen},
  \emph{\bibinfo{title}{Sound Transmission through a Fluctuating Ocean}}
  (\bibinfo{publisher}{Cambridge University Press},
  \bibinfo{address}{Cambridge}) (\bibinfo{year}{1979}).

\bibitem{Munk95}
\bibinfo{author}{W.~H. Munk}, \bibinfo{author}{P.~Worcester}, and
  \bibinfo{author}{C.~Wuncsh}, \emph{\bibinfo{title}{Ocean Acoustic
  Tomography}} (\bibinfo{publisher}{Cambridge University Press},
  \bibinfo{address}{Cambridge}) (\bibinfo{year}{1995}).

\bibitem{Worcester99}
\bibinfo{author}{P.~F. Worcester}, \bibinfo{author}{B.~D. Cornuelle},
  \bibinfo{author}{M.~A. Dzieciuch}, \bibinfo{author}{W.~H. Munk},
  \bibinfo{author}{B.~M. Howe}, \bibinfo{author}{J.~A.Mercer},
  \bibinfo{author}{R.~C. Spindel}, \bibinfo{author}{J.~A. Colosi},
  \bibinfo{author}{K.~Metzger}, \bibinfo{author}{T.~Birdsall}, and
  \bibinfo{author}{A.~B. Baggeroer}, \enquote{\bibinfo{title}{A test of
  basin-scale acoustic thermometry using a large- aperture vertical array at
  3250-km range in the eastern north pacific ocean}},
  \bibinfo{journal}{J.~Acoust.~Soc.~Am.} \textbf{\bibinfo{volume}{105}},
  \bibinfo{pages}{3185--3201} (\bibinfo{year}{1999}).

\bibitem{Colosi99}
\bibinfo{author}{J.~A. Colosi}, \bibinfo{author}{E.~K. Scheer},
  \bibinfo{author}{S.~M. Flatt\'{e}}, \bibinfo{author}{B.~D. Cornuelle},
  \bibinfo{author}{M.~A. Dzieciuch}, \bibinfo{author}{W.~H. Munk},
  \bibinfo{author}{P.~F. Worcester}, \bibinfo{author}{B.~M. Howe},
  \bibinfo{author}{J.~A.Mercer}, \bibinfo{author}{R.~C. Spindel},
  \bibinfo{author}{K.~Metzger}, \bibinfo{author}{T.~Birdsall}, and
  \bibinfo{author}{A.~B. Baggeroer}, \enquote{\bibinfo{title}{Comparisons of
  measured and predicted acoustic fluctuations for a 3250-km propagation
  experiment in the eastern north pacific ocean}},
  \bibinfo{journal}{J.~Acoust.~Soc.~Am.} \textbf{\bibinfo{volume}{105}},
  \bibinfo{pages}{3202--3218} (\bibinfo{year}{1999}).

\bibitem{Makarov09}
\bibinfo{author}{D.~Makarov}, \bibinfo{author}{S.~Prants},
  \bibinfo{author}{A.~Virovlyansky}, and \bibinfo{author}{G.~M. Zaslavksy},
  \emph{\bibinfo{title}{Ray and wave chaos in ocean acoustics: chaos in
  waveguides}}, volume \bibinfo{volume}{Vol.~1} of
  \emph{\bibinfo{series}{Complexity, Nonlinearity and Chaos}}
  (\bibinfo{publisher}{World Scientific Publishing Company})
  (\bibinfo{year}{2009}).

\bibitem{Simmen97}
\bibinfo{author}{J.~Simmen}, \bibinfo{author}{S.~M. Flatt\'{e}}, and
  \bibinfo{author}{G.-Y. Wang}, \enquote{\bibinfo{title}{Wavefront folding,
  chaos, and diffraction for sound propagation through ocean internal waves}},
  \bibinfo{journal}{J.~Acoust.~Soc.~Am.} \textbf{\bibinfo{volume}{102}},
  \bibinfo{pages}{239--255} (\bibinfo{year}{1997}).

\bibitem{Brown03}
\bibinfo{author}{M.~G. Brown}, \bibinfo{author}{J.~A. Colosi},
  \bibinfo{author}{S.~Tomsovic}, \bibinfo{author}{A.~L. Virovlyansky},
  \bibinfo{author}{M.~A. Wolfson}, and \bibinfo{author}{G.~M. Zaslavsky},
  \enquote{\bibinfo{title}{Ray dynamics in long-range deep ocean sound
  propagation}}, \bibinfo{journal}{J.~Acoust.~Soc.~Am.}
  \textbf{\bibinfo{volume}{113}}, \bibinfo{pages}{2533--2547}
  (\bibinfo{year}{2003}), \bibinfo{note}{nlin.CD/0109027}.

\bibitem{Beronvera03}
\bibinfo{author}{F.~J. Beron-Vera}, \bibinfo{author}{M.~G. Brown},
  \bibinfo{author}{J.~A. Colosi}, \bibinfo{author}{S.~Tomsovic},
  \bibinfo{author}{A.~L. Virovlyansky}, \bibinfo{author}{M.~A. Wolfson}, and
  \bibinfo{author}{G.~M. Zaslavsky}, \enquote{\bibinfo{title}{Ray dynamics in a
  long-range acoustic propagation experiment}},
  \bibinfo{journal}{J.~Acoust.~Soc.~Am.} \textbf{\bibinfo{volume}{114}},
  \bibinfo{pages}{1226--1242} (\bibinfo{year}{2003}),
  \bibinfo{note}{arXiv:0301026 [nlin.CD]}.

\bibitem{Tomsovic10}
\bibinfo{author}{S.~Tomsovic} and \bibinfo{author}{M.~G. Brown},
  \enquote{\bibinfo{title}{Ocean acoustics: a novel laboratory for wave
  chaos}}, in \emph{\bibinfo{booktitle}{New directions in linear acoustics and
  vibration: random matrix theory, quantum chaos and complexity}}, edited by
  \bibinfo{editor}{R.~Weaver} and \bibinfo{editor}{M.~Wright},
  \bibinfo{pages}{169--187} (\bibinfo{publisher}{Cambridge University Press},
  \bibinfo{address}{New York}) (\bibinfo{year}{2010}).

\bibitem{Tappert77}
\bibinfo{author}{F.~D. Tappert}, \enquote{\bibinfo{title}{The parabolic
  approximation method}}, in \emph{\bibinfo{booktitle}{Lecture Notes in
  Physics, Vol.~{\bf 70}, Wave propagation and underwater acoustics}}, edited
  by \bibinfo{editor}{J.~B. Keller} and \bibinfo{editor}{J.~S. Papadakis},
  \bibinfo{pages}{224--287} (\bibinfo{publisher}{Springer-Verlag},
  \bibinfo{address}{New York}) (\bibinfo{year}{1977}).

\bibitem{Palmer88}
\bibinfo{author}{D.~R. Palmer}, \bibinfo{author}{M.~G. Brown},
  \bibinfo{author}{F.~D. Tappert}, and \bibinfo{author}{H.~F. Bezdek},
  \enquote{\bibinfo{title}{Classical chaos in nonseparable wave propagation
  problems}}, \bibinfo{journal}{Geophys.~Res.~Lett.}
  \textbf{\bibinfo{volume}{15}}, \bibinfo{pages}{569--572}
  (\bibinfo{year}{1988}).

\bibitem{Smith92a}
\bibinfo{author}{K.~B. Smith}, \bibinfo{author}{M.~G. Brown}, and
  \bibinfo{author}{F.~D. Tappert}, \enquote{\bibinfo{title}{Ray chaos in
  underwater acoustics}}, \bibinfo{journal}{J.~Acoust.~Soc.~Am.}
  \textbf{\bibinfo{volume}{91}}, \bibinfo{pages}{1939--1949}
  (\bibinfo{year}{1992}).

\bibitem{Smith92b}
\bibinfo{author}{K.~B. Smith}, \bibinfo{author}{M.~G. Brown}, and
  \bibinfo{author}{F.~D. Tappert}, \enquote{\bibinfo{title}{Acoustic ray chaos
  induced by mesocsale ocean structure}},
  \bibinfo{journal}{J.~Acoust.~Soc.~Am.} \textbf{\bibinfo{volume}{91}},
  \bibinfo{pages}{1950--1959} (\bibinfo{year}{1992}).

\bibitem{Beenakker97RMP}
\bibinfo{author}{C.~W.~J. Beenakker}, \enquote{\bibinfo{title}{Random-matrix
  theory of quantum transport}}, \bibinfo{journal}{Rev.~Mod.~Phys.}
  \textbf{\bibinfo{volume}{69}}, \bibinfo{pages}{731--808}
  (\bibinfo{year}{1997}).

\bibitem{Marcus92}
\bibinfo{author}{C.~M. Marcus}, \bibinfo{author}{A.~J. Rimberg},
  \bibinfo{author}{R.~M. Westervelt}, \bibinfo{author}{P.~F. Hopkins}, and
  \bibinfo{author}{A.~C. Gossard}, \enquote{\bibinfo{title}{Conductance
  fluctuations and chaotic scattering in ballistic microstructures}},
  \bibinfo{journal}{Phys.~Rev.~Lett.} \textbf{\bibinfo{volume}{69}},
  \bibinfo{pages}{506} (\bibinfo{year}{1992}).

\bibitem{Topinka00}
\bibinfo{author}{M.~A. Topinka}, \bibinfo{author}{B.~J. LeRoy},
  \bibinfo{author}{S.~E.~J. Shaw}, \bibinfo{author}{E.~J. Heller},
  \bibinfo{author}{R.~M. Westervelt}, \bibinfo{author}{K.~D. Maranowski}, and
  \bibinfo{author}{A.~C. Gossard}, \enquote{\bibinfo{title}{Imaging coherent
  electron flow from a quantum point contact}}, \bibinfo{journal}{Science}
  \textbf{\bibinfo{volume}{289}}, \bibinfo{pages}{2323--2326}
  (\bibinfo{year}{2000}).

\bibitem{Topinka01}
\bibinfo{author}{M.~A. Topinka}, \bibinfo{author}{B.~J. LeRoy},
  \bibinfo{author}{R.~M. Westervelt}, \bibinfo{author}{S.~E.~J. Shaw},
  \bibinfo{author}{R.~Fleischmann}, \bibinfo{author}{E.~J. Heller},
  \bibinfo{author}{K.~D. Maranowski}, and \bibinfo{author}{A.~C. Gossard},
  \enquote{\bibinfo{title}{Coherent branched flow in a two-dimensional electron
  gas}}, \bibinfo{journal}{Nature} \textbf{\bibinfo{volume}{410}},
  \bibinfo{pages}{183--186} (\bibinfo{year}{2001}).

\bibitem{Wolfson01}
\bibinfo{author}{M.~A. Wolfson} and \bibinfo{author}{S.~Tomsovic},
  \enquote{\bibinfo{title}{On the stability of long-range sound propagation
  through a structured ocean}}, \bibinfo{journal}{J.~Acoust.~Soc.~Am.}
  \textbf{\bibinfo{volume}{109}}, \bibinfo{pages}{2693--2703}
  (\bibinfo{year}{2001}), \bibinfo{note}{arXiv:0002030 [nlin.CD]}.

\bibitem{Hegewisch12}
\bibinfo{author}{K.~C. Hegewisch} and \bibinfo{author}{S.~Tomsovic},
  \enquote{\bibinfo{title}{Random matrix theory for underwater sound
  propagation}}, \bibinfo{journal}{Europhys. ~Lett.}
  \textbf{\bibinfo{volume}{97}}, \bibinfo{pages}{34002} (\bibinfo{year}{2012}),
  \bibinfo{note}{arXiv:1104.3975 [physics.ao-ph]}.

\bibitem{Hegewisch10}
\bibinfo{author}{K.~C. Hegewisch}, Ph.D. thesis, \bibinfo{school}{Washington
  State University} (\bibinfo{year}{2010}).

\bibitem{Weaver89}
\bibinfo{author}{R.~L. Weaver}, \enquote{\bibinfo{title}{Spectral statistics in
  elastodynamics}}, \bibinfo{journal}{J.~Acoust.~Soc.~Am.}
  \textbf{\bibinfo{volume}{85}}, \bibinfo{pages}{1005--1013}
  (\bibinfo{year}{1989}).

\bibitem{Wigner55}
\bibinfo{author}{E.~Wigner}, \enquote{\bibinfo{title}{Characteristic vectors of
  bordered matrices with infinite dimensions}}, \bibinfo{journal}{Ann.~of
  Math.} \textbf{\bibinfo{volume}{62}}, \bibinfo{pages}{548--564}
  (\bibinfo{year}{1955}).

\bibitem{Dyson62e}
\bibinfo{author}{F.~J. Dyson}, \enquote{\bibinfo{title}{The threefold way.
  algebraic structure of symmetry groups and ensembles in quantum mechanics}},
  \bibinfo{journal}{J.~Math.~Phys.} \textbf{\bibinfo{volume}{3}},
  \bibinfo{pages}{1199--1215} (\bibinfo{year}{1962}).

\bibitem{Dozier78}
\bibinfo{author}{L.~B. Dozier} and \bibinfo{author}{F.~D. Tappert},
  \enquote{\bibinfo{title}{Statistics of normal mode amplitudes in a random
  ocean. i. theory}}, \bibinfo{journal}{J.~Acoust.~Soc.~Am.}
  \textbf{\bibinfo{volume}{63}}, \bibinfo{pages}{353--365}
  (\bibinfo{year}{1978}).

\bibitem{Dozier78b}
\bibinfo{author}{L.~B. Dozier} and \bibinfo{author}{F.~D. Tappert},
  \enquote{\bibinfo{title}{Statistics of normal mode amplitudes in a random
  ocean. ii. computations}}, \bibinfo{journal}{J.~Acoust.~Soc.~Am.}
  \textbf{\bibinfo{volume}{63}}, \bibinfo{pages}{533--547}
  (\bibinfo{year}{1978}).

\bibitem{Morozov07}
\bibinfo{author}{A.~K. Morozov} and \bibinfo{author}{J.~A. Colosi},
  \enquote{\bibinfo{title}{Stochastic differential equation analysis for sound
  scattering by random internal waves in the ocean}},
  \bibinfo{journal}{Acoustical Physics} \textbf{\bibinfo{volume}{53}},
  \bibinfo{pages}{335--347} (\bibinfo{year}{2007}).

\bibitem{Colosi09}
\bibinfo{author}{J.~A. Colosi} and \bibinfo{author}{A.~K. Morozov},
  \enquote{\bibinfo{title}{Statistics of normal mode amplitudes in an ocean
  with random sound-speed perturbations: cross-mode coherence and mean
  intensity}}, \bibinfo{journal}{J.~Acoust.~Soc.~Am.}
  \textbf{\bibinfo{volume}{126}}, \bibinfo{pages}{1026--1035}
  (\bibinfo{year}{2009}).

\bibitem{Mello10}
\bibinfo{author}{P.~A. Mello} and \bibinfo{author}{N.~Kumar},
  \emph{\bibinfo{title}{Quantum transport in mesoscopic systems. Complexity and
  Statistical Fluctuations}} (\bibinfo{publisher}{Oxford University Press},
  \bibinfo{address}{Oxford}) (\bibinfo{year}{2010}).

\bibitem{Munk74}
\bibinfo{author}{W.~H. Munk}, \enquote{\bibinfo{title}{Sound channel in an
  exponentially stratified ocean with applications to sofar}},
  \bibinfo{journal}{J.~Acoust.~Soc.~Am.} \textbf{\bibinfo{volume}{55}},
  \bibinfo{pages}{220--226} (\bibinfo{year}{1974}).

\bibitem{Colosi98}
\bibinfo{author}{J.~A. Colosi} and \bibinfo{author}{M.~G. Brown},
  \enquote{\bibinfo{title}{Efficient numerical simulation of stochastic
  internal-wave induced sound-speed perturbation fields}},
  \bibinfo{journal}{J.~Acoust.~Soc.~Am.} \textbf{\bibinfo{volume}{103}},
  \bibinfo{pages}{2232--2235} (\bibinfo{year}{1998}).

\bibitem{Hegewisch05}
\bibinfo{author}{K.~C. Hegewisch}, \bibinfo{author}{N.~R. Cerruti}, and
  \bibinfo{author}{S.~Tomsovic}, \enquote{\bibinfo{title}{Ocean acoustic wave
  propagation and ray method correspondence: internal waves}},
  \bibinfo{journal}{J.~Acoust.~Soc.~Am.} \textbf{\bibinfo{volume}{117}},
  \bibinfo{pages}{1582--1594} (\bibinfo{year}{2005}),
  \bibinfo{note}{arXiv:0312150 [physics]}.

\bibitem{Voronovich06}
\bibinfo{author}{A.~G. Voronovich} and \bibinfo{author}{V.~E. Ostashev},
  \enquote{\bibinfo{title}{Low-frequency sound scattering by internal waves in
  the ocean}}, \bibinfo{journal}{J.~Acoust.~Soc.~Am.}
  \textbf{\bibinfo{volume}{119}}, \bibinfo{pages}{1406--1419}
  (\bibinfo{year}{2006}).

\bibitem{Voronovich09}
\bibinfo{author}{A.~G. Voronovich} and \bibinfo{author}{V.~E. Ostashev},
  \enquote{\bibinfo{title}{Coherence function of a sound field in an oceanic
  waveguide with horizontally isotropic statistics}},
  \bibinfo{journal}{J.~Acoust.~Soc.~Am.} \textbf{\bibinfo{volume}{125}},
  \bibinfo{pages}{99--110} (\bibinfo{year}{2009}).

\bibitem{PorterBook}
\bibinfo{author}{C.~E. Porter}, \emph{\bibinfo{title}{Statistical Theories of
  Spectra: Fluctuations}} (\bibinfo{publisher}{Academic Press},
  \bibinfo{address}{New York}) (\bibinfo{year}{1965}).

\bibitem{Pandey79}
\bibinfo{author}{A.~Pandey}, \enquote{\bibinfo{title}{Statistical properties of
  many-particle spectra: Iii. ergodic behavior in random-matrix ensembles}},
  \bibinfo{journal}{ap} \textbf{\bibinfo{volume}{119}},
  \bibinfo{pages}{170--191} (\bibinfo{year}{1979}).

\bibitem{Perez07}
\bibinfo{author}{L.~S. Froufe-Perez}, \bibinfo{author}{M.~Yepez},
  \bibinfo{author}{P.~A. Mello}, and \bibinfo{author}{J.~J. Saenz},
  \enquote{\bibinfo{title}{Statistical scattering of waves in disordered
  waveguides: From microscopic potentials to limiting macroscopic statistics}},
  \bibinfo{journal}{Phys.~Rev.~E} \textbf{\bibinfo{volume}{75}},
  \bibinfo{pages}{031113} (\bibinfo{year}{2007}).

\bibitem{Chernov75}
\bibinfo{author}{L.~A. Chernov}, \emph{\bibinfo{title}{Waves in
  Randomly-Inhomogeneous Media}} (\bibinfo{publisher}{Nauka},
  \bibinfo{address}{Moscow}) (\bibinfo{year}{1975}), \bibinfo{note}{(in
  Russian)}.

\bibitem{Flatte83}
\bibinfo{author}{S.~M. Flatt\'e}, \enquote{\bibinfo{title}{Wave propagation
  through random media: Contributions from ocean acoustics}},
  \bibinfo{journal}{Proc.~IEEE} \textbf{\bibinfo{volume}{71}},
  \bibinfo{pages}{1267--1294} (\bibinfo{year}{1983}).

\bibitem{uffelen09}
\bibinfo{author}{L.~J. van Uffelen}, \bibinfo{author}{P.~F. Worcester},
  \bibinfo{author}{M.~A. Dzieciuch}, and \bibinfo{author}{D.~L. Rudnick},
  \enquote{\bibinfo{title}{The vertical structure of shadow-zone arrivals at
  long range in the ocean}}, \bibinfo{journal}{J.~Acoust.~Soc.~Am.}
  \textbf{\bibinfo{volume}{125}}, \bibinfo{pages}{3569--3588}
  (\bibinfo{year}{2009}).

\bibitem{uffelen10}
\bibinfo{author}{L.~J. van Uffelen}, \bibinfo{author}{P.~F. Worcester},
  \bibinfo{author}{M.~A. Dzieciuch}, \bibinfo{author}{D.~L. Rudnick}, and
  \bibinfo{author}{J.~A. Colosi}, \enquote{\bibinfo{title}{Effects of upper
  ocean sound-speed structure on deep acoustic shadow-zone arrivals at 500- and
  1000-km range}}, \bibinfo{journal}{J.~Acoust.~Soc.~Am.}
  \textbf{\bibinfo{volume}{127}}, \bibinfo{pages}{2169--2181}
  (\bibinfo{year}{2010}).

\bibitem{Garrett79}
\bibinfo{author}{C.~J.~R. Garrett} and \bibinfo{author}{W.~H. Munk},
  \enquote{\bibinfo{title}{Internal waves in the ocean}},
  \bibinfo{journal}{Annu.~Rev.~Fluid Mech.} \textbf{\bibinfo{volume}{11}},
  \bibinfo{pages}{339--369} (\bibinfo{year}{1979}).

\end{thebibliography}

\end{document}